\documentclass[a4paper,11pt]{article}
\usepackage[T1]{fontenc}
\usepackage[utf8]{inputenc}
\usepackage{setspace}
\usepackage[margin=1.1in]{geometry}
\usepackage{float}
\usepackage{amsmath}
\usepackage{textcomp}
\usepackage{subfig}
\usepackage{caption}
\usepackage{graphics,epsfig,psfig}
\usepackage[normalem]{ulem}
\usepackage[dvipsnames]{color}
\usepackage[]{inputenc,amssymb}

\title{\textbf {STUDY OF REDSHIFTED HI FROM THE EPOCH OF REIONIZATION  WITH DRIFT SCAN}}
\date{} 
\author
{Sourabh Paul$^{1}$ \thanks{sourabh@rri.res.in}, Shiv K. Sethi$^1$
\thanks{sethi@rri.res.in}, Ravi Subrahmanyan$^{1,11}$, N. Udaya Shankar$^1$, \\
K. S. Dwarakanath$^1$, Avinash A. Deshpande$^1$, Gianni Bernardi$^3$, Judd D. Bowman$^5$,\\
Frank Briggs$^{6,11}$, Roger J. Cappallo$^8$, Brian E. Corey$^8$, David Emrich$^2$,\\
Bryan M. Gaensler$^{9,11}$, Robert F. Goeke$^{8}$, Lincoln J. Greenhill$^4$, Bryna J. Hazelton$^{12}$,\\
Jacqueline N. Hewitt$^{10}$, Melanie Johnston-Hollitt$^{13}$, David L. Kaplan$^{14}$,\\
Justin C. Kasper$^4$, Eric Kratzenberg$^8$, Colin J. Lonsdale$^8$, Mervyn J. Lynch$^2$,\\
S. Russell McWhirter$^8$, Daniel A. Mitchell$^{7,11}$, Miguel F. Morales$^{12}$, Edward H. Morgan$^{10}$,\\
Divya Oberoi$^{15}$, Stephen M. Ord$^{2,11}$, Thiagaraj Prabu$^1$, Alan E. E. Rogers$^8$,\\
Anish A. Roshi$^{16}$, K. S. Srivani$^1$, Steven J. Tingay$^{2,11}$, Randall B. Wayth$^{2,11}$,\\
Mark Waterson$^{2,6}$, Rachel L. Webster$^{11,17}$, Alan R. Whitney$^8$,\\
Andrew J. Williams$^2$, Christopher L. Williams$^{10}$\\
\\
$^1${\footnotesize Raman Research Institute, Bangalore, India}\\
$^2${\footnotesize Curtin University, Perth, Australia}\\
$^3${\footnotesize Square Kilometre Array South Africa (SKA SA)}\\
$^4${\footnotesize Harvard-Smithsonian Center for Astrophysics, Cambridge, USA}\\
$^5${\footnotesize Arizona State University, Tempe, USA}\\
$^6${\footnotesize The Australian National University, Canberra, Australia}\\
$^7${\footnotesize CSIRO Astronomy and Space Science, Australia}\\
$^8${\footnotesize MIT Haystack Observatory, Westford, USA}\\
$^9${\footnotesize University of Sydney}\\
$^{10}${\footnotesize MIT Kavli Institute for Astrophysics and Space Research, Cambridge, USA}\\
$^{11}${\footnotesize ARC Centre of Excellence for All-sky Astrophysics (CAASTRO)}\\
$^{12}${\footnotesize University of Washington, Seattle, USA}\\
$^{13}${\footnotesize Victoria University of Wellington, New Zealand}\\
$^{14}${\footnotesize University of Wisconsin-Milwaukee, Milwaukee, USA}\\
$^{15}${\footnotesize National Center for Radio Astrophysics, Pune, India}\\
$^{16}${\footnotesize Natinal Radio Astronomy Observatory}\\
$^{17}${\footnotesize The University of Melbourne, Melbourne, Australia}\\}

\begin{document}
\bibliography{ref.bib}
\maketitle

\begin{abstract}
The detection of the Epoch of  Reionization (EoR) in the
redshifted 21-cm line is a challenging task.
Here we formulate the detection of the EoR signal using the drift scan
strategy.  This method potentially
has better instrumental stability as  compared to the case where a single
 patch of sky is tracked.
We demonstrate that the correlation time between
measured visibilities could extend up to $1-2$~hr for an
interferometer array such as the Murchison Widefield Array (MWA),
which has a wide primary beam.
We estimate  the EoR power based on
cross-correlation of visibilities  across time and show that
the drift scan strategy is capable of the  
detection  of the EoR signal 
with comparable/better signal-to-noise as compared to
 the tracking case.
We also estimate the   visibility correlation for a set of  
  bright point sources 
and argue that the statistical inhomogeneity  of bright point sources
might allow their separation from the EoR signal.
\end{abstract}

\section{Introduction}
The dark age of the universe ends with the formation of first galaxies. The
 ultravilolet photons from these galaxies start ionizing the
neutral HI gas in the universe and form large ionized bubbles. Eventually
 these ionized bubbles grow  in size and merge  until
there is no neutral hydrogen left in the universe except in 
a   dense  optically-thick clouds.
 This major phase transition of the universe is marked as the 
Epoch of Reionization (EoR). Current observational constraints imply that EoR
 occurred in  the redshift range $z\simeq 6\hbox{--}15$ (Fan et al. 2006; Komatsu et al. 2010).  
\vspace{10 pt}

Apart from the CMBR anisotropy measurements which give the integrated optical
depth through the reionization epoch, the redshifted 21-cm line transition
 from neutral Hydrogen  is the other  major  probe for studying this epoch. An 
obvious advantage with the 21-cm probe  is that it could reveal the 
 three-dimensional
structure of the EoR.  Currently many radio interferometers 
are operational with the specific aim   to detect the EoR (MWA (Tingay et al. 2013, \cite{2013PASA...30....7T}), 
LOFAR (Van Haarlem, M. P. et al. 2013, \cite{LOFAR}), and PAPER (Parsons, A. R. et al. 2013, \cite{PAPER})).
However, detection of the EoR signal
 is a challenging task for
 any present day EoR experiments for multiple reasons.  First, it is 
an extremely weak signal (brightness temperature fluctuations $\Delta T_{\rm B} \simeq 10 \, \rm mk$) and only a statistical detection of the signal might
be possible with the presently operational  interferometers. Second,  in the 
frequency domain 50--250 MHz, both the galactic and extragalactic 
 foregrounds are larger than  the observed signal. 
Major contribution to  foregrounds come from the 
synchrotron emission of relativistic electrons in our Galaxy, radio galaxies, 
resolved 
supernovae remnants,  free-free emission, and the  unresolved extragalactic 
radio sources  (S. Zaroubi, 2012, \cite{2013ASSL..396...45Z}). Although 
radio interferometers focus on the fluctuations
 in the signal,  the fluctuations in  foregrounds on 
relevant angular scales  are 10-1000 times 
 higher than the desired cosmological signal. 

\vspace{10 pt}

 The  statistical detection of the EoR signal 
 requires a stable instrument and a large amount of data to
 reduce the  thermal noise, in addition to measuring  and
subtracting  the foregrounds.  The traditional tracking mode of 
observation may not be useful for this purpose as it 
 leads to a time dependent primary beam as the pointing center 
is moved. In  the drift scan technique  the pointing centre
is fixed at a particular point on the sky and the observation
 is carried out for  a variable sky pattern.
One  advantage of this technique is the stability  of the system.
 
\vspace{10 pt}

In this paper we describe a   methodology  based on
drift scans that exploits the  correlation between 
 visibilities measured at different times to estimate the EoR signal.
 In particular, our aim is to infer the efficacy of such a method 
for a wide field-of-view instrument such as MWA. 

In the next section we delineate the basic formalism. In section~3, we apply
the method to the system parameters of MWA. In section~4, we compute the 
noise on the estimator of the EoR proposed in this paper and compare the 
drift scan results with the expected noise in the tracking case.  In Section~5, 
we discuss briefly how our method might 
potentially allow foregrounds represented by bright point sources
to  be separated from the EoR signal. In section~6 we summarize our results.

\section{Visibility Correlation in Drift scan}
The basic aim of a radio interferometer is to calculate the spatial correlation of the electric fields 
from a distant source in the sky. The measured spatial correlation is termed as 'visibility' and is given by: 
\begin{equation}
 V_\nu({\bf U})=\int A(\vec{\theta}) I_\nu (\vec{\theta})e^{-i2\pi {\bf U}\cdot\vec{\theta}} d\Omega \label{visibility}
\end{equation}
 Here {\bf U} denotes the baseline vector joining a pair of antennas, measured in units of wavelength, projected on a plane 
perpendicular to the direction of observation and $\vec{\theta}$ refers to the position on the sky.
$ A(\vec{\theta})$ is the primary beam pattern and $I_\nu(\vec{\theta})$ is the observed intensity at frequency $\nu$. All the other variables 
have their usual meaning, e.g. \cite{Thompson}.

In the case of high-redshift  HI emission, the specific intensity from any direction $\vec{\theta}$  at the redshifted  frequency $\nu = 1420/(1+z) \, \rm MHz$, 
 can be decomposed into two parts:
\begin{equation}
 I_\nu(\vec{\theta})=\bar{I}_\nu + \Delta I_\nu(\vec{\theta})
\end{equation}
where $\bar{I}_\nu$ and $\Delta I_\nu(\vec{\theta})$ are the isotropic and fluctuating components of the specific intensity.

This allows us to   express  the  visibility arising from HI emission as:
\begin{equation}
 V_\nu({\bf U})=\int A(\vec{\theta}) \Delta I_\nu (\vec{\theta})e^{-i2\pi {\bf U}\cdot\vec{\theta}} d\Omega \label{vis}
\end{equation}
Here only the fluctuating component appears since the isotropic component does not contribute to the visibility. We drop the w-term in writing the relation
 between 
the visibility and specific intensity. We discuss the impact of the 
w-term  in Appendix~B. 

The  fluctuating component  of HI emission can be expressed in terms of $\Delta_{HI}({\bf k})$,the Fourier transform of the density 
contrast of the HI number density $\Delta n_{\rm HI}({\bf x})/\bar{n}_{\rm HI}$ 
(e.g. Bharadwaj \& Sethi 2001  \cite{2001JApA...22..293B}; Morales \& Hewitt 2004 \cite{2004ApJ...615....7M}):
\begin{equation}
 \Delta I_\nu(\vec{\theta})=\bar{I}_\nu\int\frac{d^3k}{(2\pi)^3}\Delta_{HI}({\bf k})
 e^{ir_\nu (k_\parallel+{\bf{k_\perp}}\cdot\vec{\theta})}
\end{equation}
Here  $k_\parallel$ and ${\bf k_\perp}$ refer to the components of 
comoving wave vector  ${\bf k}$ along line of sight and in the  plane of the sky, respectively and $r_\nu$ is the comoving  distance. With these definitions
 we can  expand the phase term  ${\bf k}.{\bf r}$, as shown in  Eq.~(4).  The 3D Fourier
transform can be understood as  performing 1D Fourier transform along
the  line of sight followed by a 2D fourier transform on the sky plane, or 
$d^3k = dk_\parallel d^2 k_\perp$. Our formulation allows us to treat the integral on the sky plane using cartesian coordinates, 
$d^3k = dk_\parallel dk_{\perp 1}dk_{\perp 2}$ (Eq.~(15) and Appendix A for details). 

\vspace{10 pt}
 Eq.~(\ref{vis}) can thus be expressed as:
\begin{equation}
 V_\nu({\bf U})=\bar{I}_\nu\int\frac{d^3k}{(2\pi)^3}\Delta_{HI}({\bf k})e^{ir_\nu k_\parallel}
 \int d\Omega A(\vec{\theta})\exp\left[-2\pi i\left({\bf U}-
 \frac{{\bf k_\perp}r_\nu}{2\pi}\right)\ldotp\vec{\theta}\right] \label{visu}
\end{equation}
The second integral over  the primary beam $A(\vec{\theta})$ can be  denoted as:
\begin{equation}
 a\left({\bf U}-\frac{{\bf k_\perp}r_\nu}{2\pi}\right)\equiv \int d\Omega A(\vec{\theta})
 \exp\left[-2\pi i\left({\bf U}-\frac{{\bf k_\perp}r_\nu}{2\pi}\right)\ldotp\vec{\theta}\right]
\end{equation}
Thus,  finally,  Eq.~(\ref{visu}) takes the form:
\begin{equation}
 V_\nu({\bf U})=\bar{I_\nu}\int\frac{d^3k}{(2\pi)^3}\Delta_{HI}({\bf k})
 e^{ir_\nu k_\parallel}a\left({\bf U}-\frac{{\bf k_\perp}r_\nu}{2\pi}\right)\label{viscorr}
\end{equation}

If the first visibility measurement is obtained at $t = 0$, then,
using Eq.~(\ref{viscorr}), the visibility measured at a later time $t$, for 
a drift scan,   can be written as:
\begin{eqnarray}
 V_\nu({\bf U},t)&=&\bar{I}_\nu\int\frac{d^3k}{(2\pi)^3}\Delta_{HI}({\bf k})
 \int d^2\theta A(\vec{\theta})\exp\left[ir_\nu\left (k_\parallel+{\bf k_\perp}.(\vec\theta-{\Delta\vec\theta}(t)) \right )\right]\exp(-2\pi i{\bf U}.\vec\theta) \nonumber \\
&=& {\bar I}_\nu\int\frac{d^3k}{(2\pi)^3}\Delta_{HI}({\bf k})e^{ir_\nu k_\parallel}
 \int d^2\theta A(\vec{\theta})\exp\left[-2\pi i\left({\bf U}-
 \frac{{\bf k_\perp}r_\nu}{2\pi}\right)\ldotp\vec{\theta}\right]\exp[-ir_\nu{\bf k_\perp}.{\Delta\vec\theta}(t)]\nonumber \\ 
\label{viscorrt}
 \end{eqnarray}
Here $\Delta{\vec\theta}(t)$ is the angular shift of the intensity pattern 
in the time period $t$. Eq.~(\ref{viscorrt}) follows from Eqs~(3)--(5) for 
a changing intensity pattern.  In a drift scan,  the phase center and the 
primary beam remain fixed and the only change in the visibility occurs
owing to the  changing intensity pattern of the sky  with respect to the phase
 center.

Our aim is  to calculate the correlation between the visibilities
 measured at two different times (separated by $t$), 
 by two  baselines ${\bf U}$ and ${\bf U'}$, and 
at   frequencies $\nu$ and $\nu'$. We note that 
  the frequency coverage   is far smaller than the central frequency: $|\nu'-\nu| << \nu$. This allows us to write: $|r_\nu'-r_\nu| \equiv 
\Delta r_\nu = r_\nu' |\nu'-\nu|$; here $r_\nu' \equiv |dr_\nu/d\nu|$. 
 
Using  Eqs.~(\ref{viscorr}) and~(\ref{viscorrt}), we can write the visibility correlation function  as:
\begin{eqnarray}
 \langle V_\nu({\bf U})V^*_{\nu'}({\bf U'},t)\rangle &=&\bar{I_\nu}^2
\int \frac{d^3k}{(2\pi)^3}P_{HI}(k)e^{ik_\parallel \Delta r_\nu}a\left({\bf U}-\frac{{\bf k_\perp}r_\nu}{2\pi}\right)\int d^2\theta A(\vec{\theta}) \nonumber
\\
&&\:\times 
\exp\left[-2\pi i\left({\bf U'}-
 \frac{{\bf k_\perp}r_\nu}{2\pi}\right)\ldotp\vec{\theta}\right]\exp[-ir_\nu{\bf k_\perp}.\Delta\vec\theta]\nonumber \\
 \label{crosscor}
\end{eqnarray}
where  $P_{HI}({\bf k})$ is the power spectrum of the fluctuations in the HI distribution:
\begin{equation}
 \langle\Delta_{\rm HI}^*({\bf k})\Delta_{\rm HI}({\bf k'})\rangle=(2\pi)^3\delta^3({\bf k}-{\bf k'})P_{\rm HI}({\bf k})
\label{powspec3d}
\end{equation}
where $\delta^3(x)$ is the Dirac delta function and the angular bracket denotes ensemble average. The delta function captures the information that 
the HI signal is statistically homogeneous. In the usual case
of tracking a fixed region, the ensemble average $\langle \ldots \rangle$ (LHS of Eq.~(\ref{powspec3d})) to compute the power spectrum  is done by  
averaging over all modes ${\bf k}$  for a given  $|{\bf k}|$. The drift scan strategy enables another possible method to compute the power 
spectrum for modes in the plane of the sky ${\bf k_\perp}$: averaging over time for a given fixed time difference, $\Delta t$, for visibility measurements. We discuss this issue in detail in section~(4). For a statistically homogeneous 
signal, e.g. the EoR, these two methods  yield the same estimate of the 
power spectrum. However, when the
assumption of statistical homogeneity breaks down, e.g. for sparsely distributed point sources, the two methods result in different outcomes. We explicitly 
make use of this difference in our discussion of point sources in a
 section~5.
\vspace{10 pt}

The brightness temperature fluctuations $\Delta_{\rm HI}({\bf k})$ are a combination of different
physical effects, e.g. density fluctuations, ionization inhomogeneities, 
and density-ionization fraction cross-correlation, e.g. 
(Furlanetto et al. 2006, \cite{2006PhR...433..181F}; Zaldarriaga et al. 2004, \cite{2004ApJ...608..622Z}):
\begin{equation}
 \Delta_{HI}=\beta\delta_b+\beta_x\delta_x+\beta_\alpha\delta_\alpha+\beta_T\delta_T-\delta_{\partial v}
\end{equation}
Here each term refers to to the fractional variation in a particular quantity. Thus $\delta_b$ stands for variation in baryonic density,
$\delta_\alpha$ for the Ly$\alpha$ coupling coefficient $x_\alpha$, $\delta_x$ for the neutral fraction, $\delta_T$ for $T_K$, and 
$\delta_{\partial v}$ for the line of sight peculiar velocity gradient. $\beta$ factors are the corresponding expansion coefficients. For
further details  refer to Furlanetto et al. 2006, \cite{2006PhR...433..181F}.
Throughout this paper, we adopt 
 the theoretical spherically averaged power spectrum (Eq.~(\ref{powspec3d})) 
from Beardsley et al. 2013 (Figure~4 of their paper), 
\cite{2013MNRAS.429L...5B}  
(see also Furlanetto et al. 2006, \cite{2006PhR...433..181F}).

The isotropic part of the emission can be calculated as:
\begin{equation}
 \bar{I_\nu}=\frac{A_{21}\hspace{1 pt}h_P \hspace{1 pt} c \hspace{1 pt} \bar{n}_{HI}(z)}{4\pi H(z)}
\end{equation}
Here $A_{21}$ is the Einstein coefficient of the 21 cm HI transition, $h_P$ is the Planck constant, c is the speed of light
and $H(z)$ is the Hubble parameter:
\begin{equation}
 H(z)=H_0\left[\Omega_{m0}(1+z)^3+\Omega_{\Lambda 0}\right]^{1/2}
\end{equation}
$H_0$ is the value of Hubble constant at present epoch:$H_0=70$ km s$^{-1}$ Mpc$^{-1}$ and all the results are calculated using $\Omega_{m0}=0.3$ 
and $\Omega_{\Lambda 0}=0.7$ (\cite{Spergel,planck}).

\section{Drift scan visibility correlation:  MWA}
We assume the  MWA primary beam to compute the visibility correlations (Eq.~(\ref{crosscor})); MWA primary beam can be expressed as:
\begin{equation}
 A(l,m)=\frac{\sin(\pi L_x l)}{\pi L_x l}\frac{\sin(\pi L_y m)}{\pi L_y m}
\label{pribeam}
\end{equation}
Here $L_x$ and $L_y$ are sides of an aperture of an MWA tile in units of wavelength with $L_x\approx L_y\approx2$ and (l,m) are coordinates defined on the sky.

We note that for a dipole array such as MWA, Eq.~(\ref{pribeam}) is valid for 
only a phase center at the zenith. If  the phase center is changed  (e.g. for
 tracking a region), the projected area of the tile decreases which results in
an dilation of primary beam depending on the angular position of the 
phase center. We neglect this change in the paper and throughout present results
for the primary beam given by Eq.~(\ref{pribeam}). This assumption alters  
the signal, the computation of the signal-to-noise and also the impact of the 
w-term, but doesn't change our main results.  We discuss the implications of
 this assumption in Appendix~B.

The knowledge  of HI power spectrum (Eq.~(\ref{powspec3d})) and the primary beam (Eq.~(\ref{pribeam}))  allows us to compute the evolution of visibility correlations. A detailed
formulation of the sky  coordinate system  for analysing drift scans
 from any arbitrary location of an  observatory  is discussed in appendix A. 
We first discuss the fiducial case of a zenith drift  scan for an observatory located
at the the latitude $\phi$. 
The visibility correlation function for this case 
is derived in appendix A and is given by equation (\ref{vvlmn}):
\begin{eqnarray}
  \langle V_\nu({\bf U})V^*_{\nu'}({\bf U'},t)\rangle &=& \bar{I_\nu}^2\int \frac{d^3k}{(2\pi)^3}P_{HI}(k)e^{ik_\parallel \Delta r_\nu}
  \exp\left(-ir_{\nu'} k_{\perp 1}\cos\phi dH\right) \nonumber \\
  &&\:a\left[\left(u-\frac{r_\nu k_{\perp 1}}{2\pi}\right),\left(v-\frac{r_\nu k_{\perp 2}}{2\pi}\right)\right] \nonumber \\
  &&\:a\left[\left(u'-\frac{r_\nu}{2\pi}(k_{\perp 1}+k_{\perp 2}\sin\phi dH)\right)
 ,\left(v'-\frac{r_\nu}{2\pi}(k_{\perp 2}-k_{\perp 1}\sin\phi dH)\right)\right] \nonumber \\ 
\label{viscorrfin1}
 \end{eqnarray}
Here dH is the change in hour angle corresponding to time difference t; u and v are the components of the baseline vector: {\bf U}$=u{\hat u}+
v{\hat v}$.

Many generic results follow from Eq.~(\ref{viscorrfin1}) and they are common
to both tracking and drift scan cases, so we first consider 
$dH = 0$: (a)   the contribution in 
each visibility correlation   from different modes
 is significant  when 
 $k_\perp=2\pi{\bf U}/{r_\nu}\pm 1/(\theta_0 r_\nu)$, where 
$\theta_0$ is  the angular extent of the primary beam.  In other 
words, unless the two baselines being correlated  satisfy this condition 
the visibilities get decorrelated. For MWA primary beam, this implies 
 ${\bf U} - {\bf U'} \gtrsim 0.5$, (b) If the two visibilities 
being correlated are separated by  a non-zero frequency difference
 $|\nu'-\nu|$, the 
signal strength is reduced. We later show that the 
frequency difference for which the signal drops to half its 
value:  $ |\nu'-\nu|\simeq 0.5 \, \rm MHz$. 
 We note here that we assume each 
visibility measurement to have zero channel width $\Delta \nu = 0$. 
This is justified because the channel width of MWA $\Delta \nu \simeq \, \rm 40 kHz$
which is much smaller than the decorrelation width, or 
 $\Delta \nu \ll |\nu'-\nu|$ (Figure~8).

The principle aim of this 
paper is to analyse the   visibility 
decorrelation in time domain for a drifting sky.

\begin{figure}[H]
\centering
\includegraphics[width=0.9\textwidth]{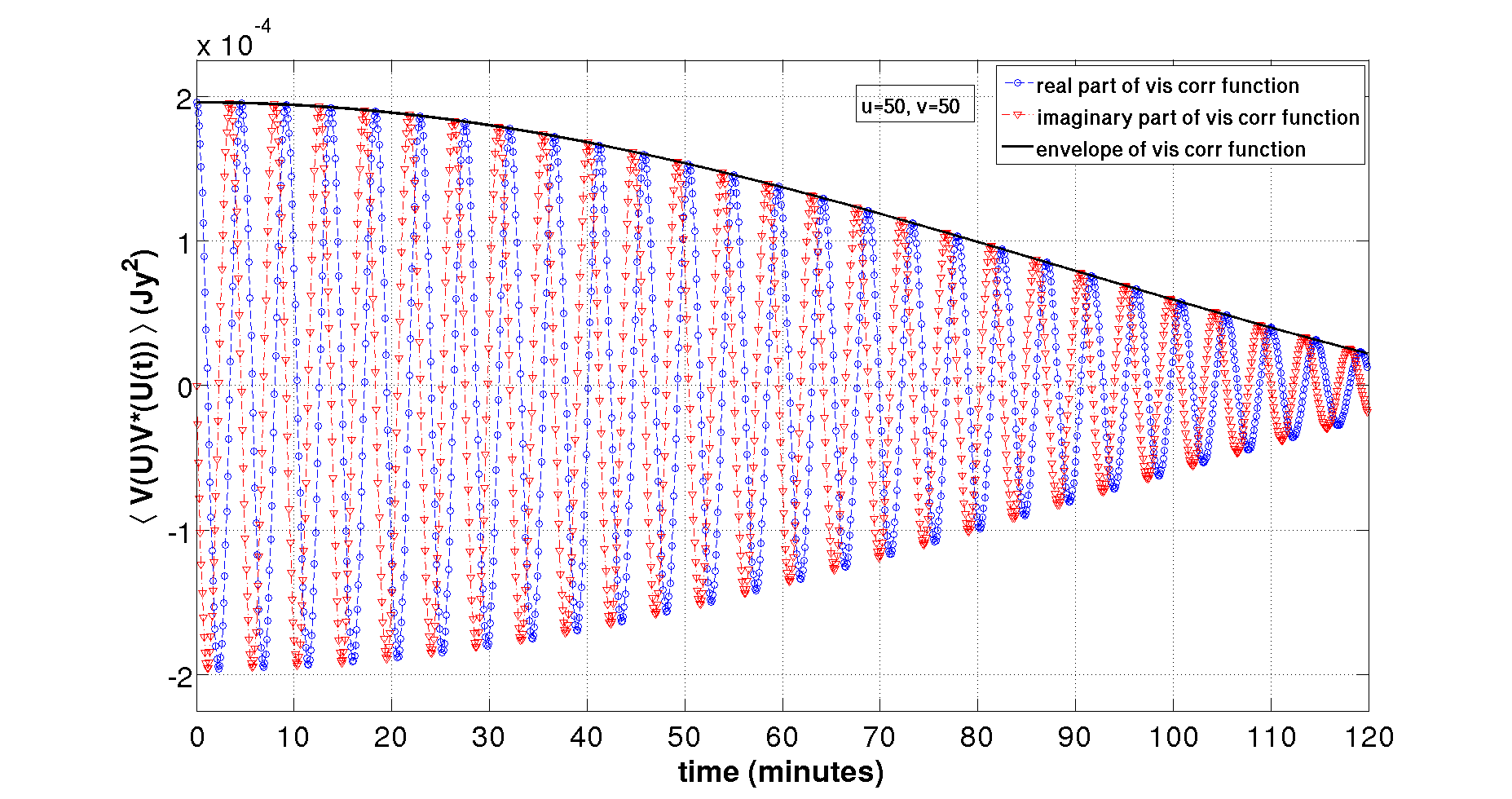}
\caption{\it Visibility Correlation function as a function of time for zenith drift from equator (latitude=0). Blue and red curves correspond
to the real and imaginary part of the visibility correlation function respectively. Black curve denotes the envelope of the Visibility correlation
function. In the figure (and all the subsequent figures that display the 
visibility correlation) the visibility correlation corresponds to the 
HI signal from EoR computed using the power spectrum of Beardsley et al. 2013 (Figure~4 of their paper), \cite{2013MNRAS.429L...5B}. The central 
frequency is assumed to be  $\nu=129$~MHz. }
\label{eq50}
\end{figure}

\begin{figure}[H]
\centering
\includegraphics[width=0.9\textwidth]{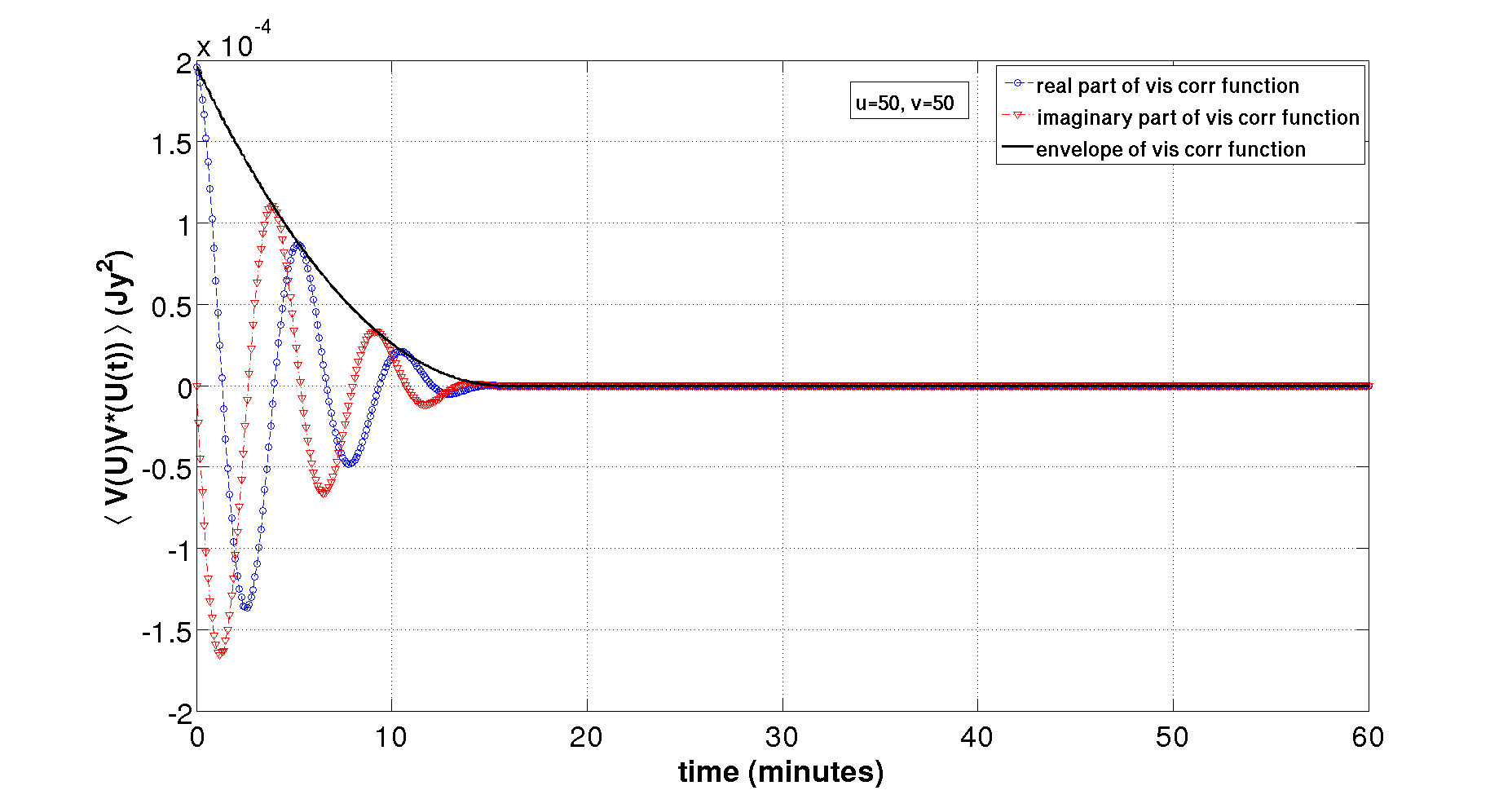}
\caption{\it Visibility Correlation function as a function of time for zenith drift from a location  with latitude \textpm$30^\circ$}
\label{30deg50}
\end{figure}

\begin{figure}[H]
\centering
\includegraphics[width=0.9\textwidth]{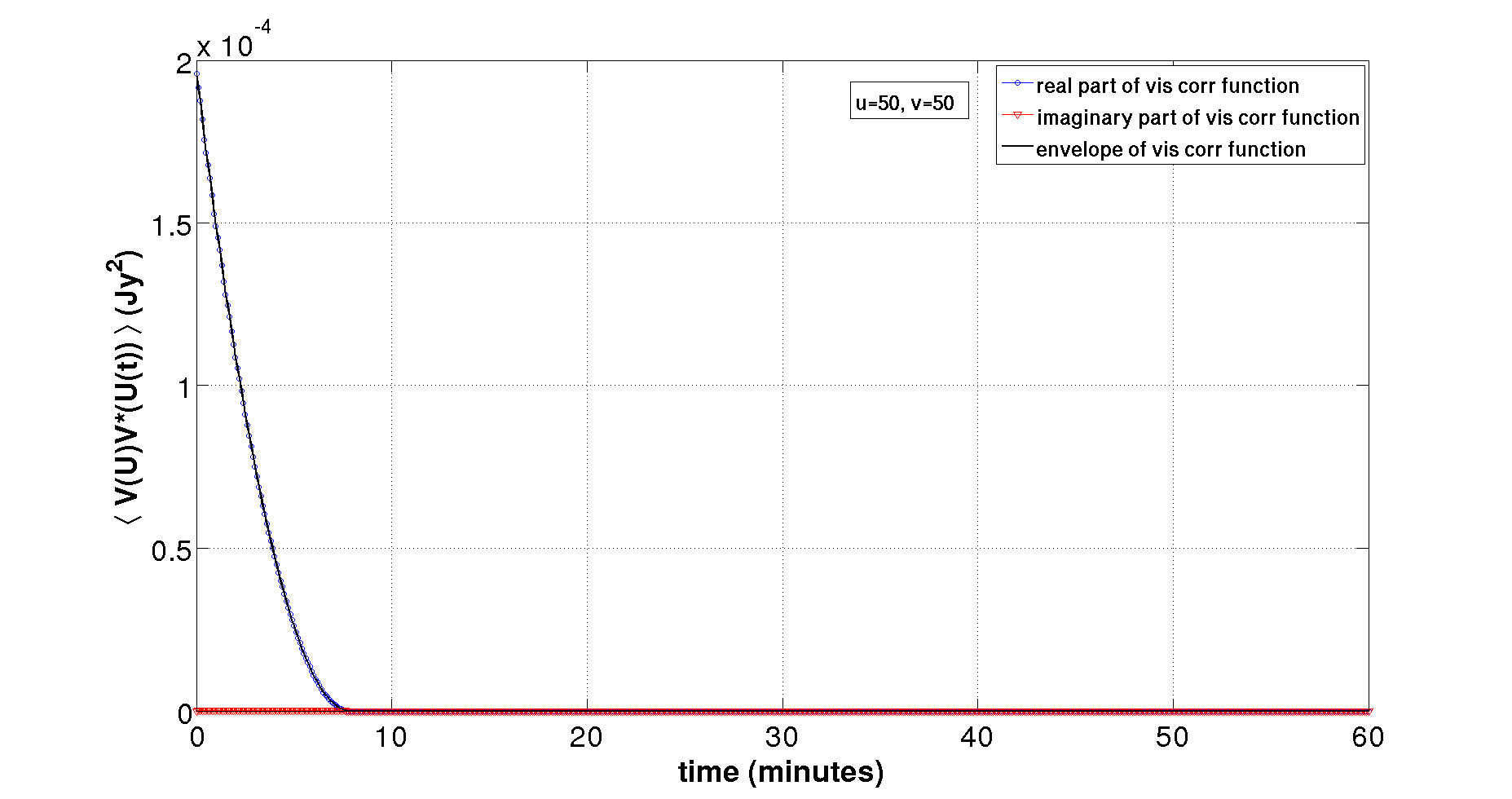}
\caption{\it Visibility Correlation function as a function of time for zenith drift from the  pole (latitude=\textpm$90^\circ$)}
\label{pole50}
\end{figure}

We show  the behaviour of visibility correlation function as a 
function of time difference for zenith drift assuming the observatory
 location to be at three different latitudes: $0^\circ$ (equator),
\textpm$30^\circ$, \textpm$90^\circ$ (pole).
The central frequency is chosen to be at $\nu=129$ MHz corresponding to redshift $z=10$ and $|\nu'-\nu|=0.0$ MHz.  
(figure \ref{freqcorr}).
The results are shown for a single
baseline vector {\bf U}=$(50,50)$  in Figures~\ref{eq50}--\ref{pole50}. The envelope of the visibility correlation function shown in the Figures is obtained by multiplying Eq.~(\ref{vvlmn}) by $\exp(-i2\pi u\cos\phi dH)$ and taking the real part of the  resulting expression. This procedure is akin to correcting for the 'shift in the phase center'.  

It is clear that at the equator the visibilities measured by the same
pair of antennas are correlated for the longest period of time. With increasing latitude of the observer 
the correlation time scale decreases and it is minimum for an observer at 
the pole (latitude 90 degree). During a drift scan, the baselines and 
primary beam remains fixed and the sources move in and out of the primary beam. 
As we show in Appendix A,   the motion of sources during a scan is 
 a combination of translation and rotation depending 
on the observer location and the field being observed. At the  equator, the 
 drift  corresponds to pure translation along east-west axis. From any other
 location there  also exists a rotational  component in the zenith drift. 
 The decorrelation time scale is shorter 
when the rotational component is present. This behaviour can be 
understood from Eq.~(\ref{viscorrfin1}). Unless  $\phi=0$, a baseline  $u$  
gets contribution from from not just $k_{\perp 1}$ but also $k_{\perp 2}$, the mode
perpendicular to  $u$  in the tracking case. A similar inference holds for 
$v$.  This results 
in decorrelation time scale much shorter than the  transit time 
of  the primary beam:   $\Delta h \simeq 1/(\sin(\phi) U \theta_0)$, $\theta_0$ is the approximate angular extent  of the primary beam. For pure translation, the 
decorrelation time scale depends only on the transit time of the primary beam.

Of the three fiducial cases we have studied (Figure~1--3),  Figure~2
is directly relevant for the location of MWA.   It is worthwhile to ask 
whether we could exploit the long time correlation of  the 
 equatorial scan using 
MWA by scanning  an equatorial
region. In Appendix A, we 
show that  if the phase center is shifted to the equatorial position (along the local meridian) then with respect to the new phase
center the drift is pure translation  and
 the decorrelation due to the rotation can be avoided for this phase center. 
For a detailed discussion see the Appendix A  and Figures~(\ref{abc}, \ref{def}).
 
For an observatory at   latitude $\phi$
the visibility correlation function with respect to the new phase center can be  written as (Eq.~(\ref{vvlmn_prime})):
\begin{eqnarray}
  \langle V_\nu({\bf U})V^*_{\nu'}({\bf U},t)\rangle &=& \bar{I_\nu}^2\int \frac{d^3k}{(2\pi)^3}P_{HI}(k)e^{ik_\parallel \Delta r_\nu}
  \exp\left(-ir_{\nu'} k_{\perp 1}\cos(\theta+\phi) dH\right) \nonumber \\
  &&\:a\left[\left(u-\frac{r_\nu k_{\perp 1}}{2\pi}\right),\left(v-\frac{r_\nu k_{\perp 2}}{2\pi}\right)\right] \nonumber \\
  &&\:a\left[\left(u-\frac{r_\nu}{2\pi}(k_{\perp 1}+k_{\perp 2}\sin(\theta+\phi) dH)\right)\right., \nonumber \\
  &&\:\left.\left(v-\frac{r_\nu}{2\pi}(k_{\perp 2}-k_{\perp 1}\sin(\theta+\phi) dH)\right)\right] \nonumber \\ 
 \end{eqnarray}
 Here $\theta$ is the angular distance of the new phase center from zenith for 
an observatory at  latitude $\phi$. To shift the phase center to the 
 equator the rotation angle is $\theta=-\phi$. For this phase center, the  
time dependence of the visibility correlation follows the behaviour
seen in Figure~1 or formally  Eq.~(\ref{vvlmn}) with $\phi = 0$ 
yields the same result as Eq.~(\ref{vvlmn_prime}) with $\phi = -\theta$. 
 In other words, the two cases---an observatory located at
 the equator performing a zenith drift scan  and 
an  observatory located at  some other latitude  scanning a region at 
the equator--are equivalent.

In Figure 5, the  time evolution of the  visibility
 correlation function  is shown for four different baselines for the equatorial
scan. 

\begin{figure}[H]
\centering
\includegraphics[width=0.9\textwidth]{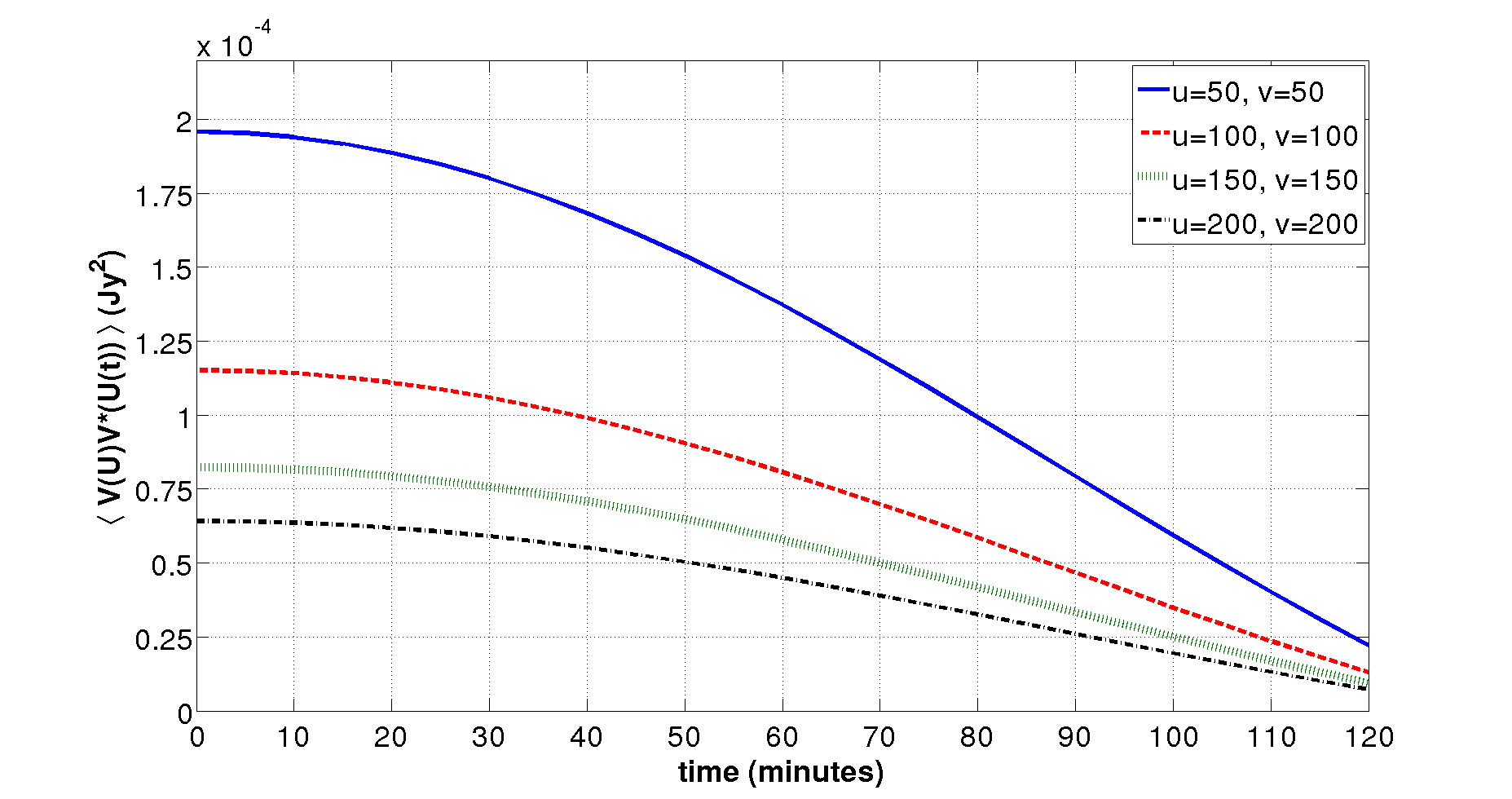}
\caption{\it Envelope of Visibility Correlation function as a function of time in drift scan mode for four different baselines. The phase center
is at zenith for an observer at the equator.}
\label{vis_t_mwa}
\end{figure}
For observing frequency $\nu=129$ MHz and $\nu' = \nu$ the visibility correlation is $\simeq 10^{-4}$Jy$^2$ for baselines $|{\bf U}|\leq 200$.
The signal strength decreases with increasing baseline length. Figure~5 also
shows that   the 
decorrelation time scale depends only  the size of the primary beam for an 
equatorial scan. 

\subsection{Correcting for Rotation}
In figures (\ref{30deg50}) and (\ref{pole50}) one sees that the rotation 
of sources in the sky plane during the drift scan reduces 
 the  time scale of decorrelation of visibility
correlation function for a given baseline.  

In a drift scan, the phase center remains fixed and therefore there is 
no change in the values of \{u,v,w\}. In other words, the set of baselines
during the scan remains the same. 

In the foregoing (Eq.~(\ref{viscorrfin1}) and the discussion
following it) we have shown that the visibilities become uncorrelated 
when ${\bf U} - {\bf U'} \gtrsim 0.5$ for the MWA primary beam. In the drift
scan case, this condition holds if  both the  visibilities are  obtained  at the
 same time. However, Eq.~(\ref{viscorrfin1}) can be used to show that 
this conclusion doesn't hold for visibilities computed at different times. 
In particular, we show that $V({\bf U},t)$ and $V({\bf U'}, t')$ can become
correlated for ${\bf U} \ne {\bf U'}$ and $t \ne t'$, if the two baselines
are related  by a special relation. We derive this relation and illustrate 
this re-correlation with  an example.

Two baselines  ${\bf U}=(u,v)$ and  ${\bf U'}=(u',v')$ can be related as:
\begin{eqnarray}
 u'=u+av+\varepsilon \nonumber \\
 v'=v-au+\varepsilon \nonumber
\end{eqnarray}
Here a and $\epsilon$ correspond to  rotation 
and translation respectively. These parameters can be solved  to give:
\begin{eqnarray}
 a=\frac{\Delta u-\Delta v}{u+v} \nonumber \\
 \varepsilon=\frac{u\Delta u+v\Delta v}{u+v} \nonumber
\end{eqnarray}
Here ${\bf \Delta U}= {\bf U} - {\bf  U'}$. 

It can be shown that for 
  two  baselines with $\varepsilon \ge 0.5$ the signal gets uncorrelated 
and cannot be re-correlated at any other time. This also means 
that two baselines with different lengths $(u^2+v^2)^{1/2}$ remain 
uncorrelated during the drift scan. However, many baselines in an experiment
such as MWA have nearly the same lengths and  are related to each other 
by a near pure  rotation denoted by the parameter $a$. We can show 
that such baselines correlate with each other  during the 
drift scan if  $a = \sin(\phi) dH$. In other words, if a  visibility is 
measured  at a time $t = 0$ for  a baseline ${\bf U}$, then this measurement
will correlate with another measurement for  a baseline ${\bf U'}$ at a time
corresponding to $dH$ if the two baselines are related by a near pure rotation
with the corresponding rotation parameter $a= \sin(\phi) dH$. We note that
this correlation can occur just once during a long scan and the time 
scale over which the baselines remain correlated   corresponds to the
  decorrelation time  for a given $|{\bf U}|$.

We illustrate this re-correlation  for a baseline ${\bf U}=(35,10)$. The other baseline ${\bf U'}= (34.8,10.7)$ corresponds to  parameters $a=-0.02$, $\varepsilon=0$.
The visibility correlation function is shown as a function of time in Figure~{\ref{VV'}.
\begin{figure}[H]
\centering
\includegraphics[width=0.8\textwidth]{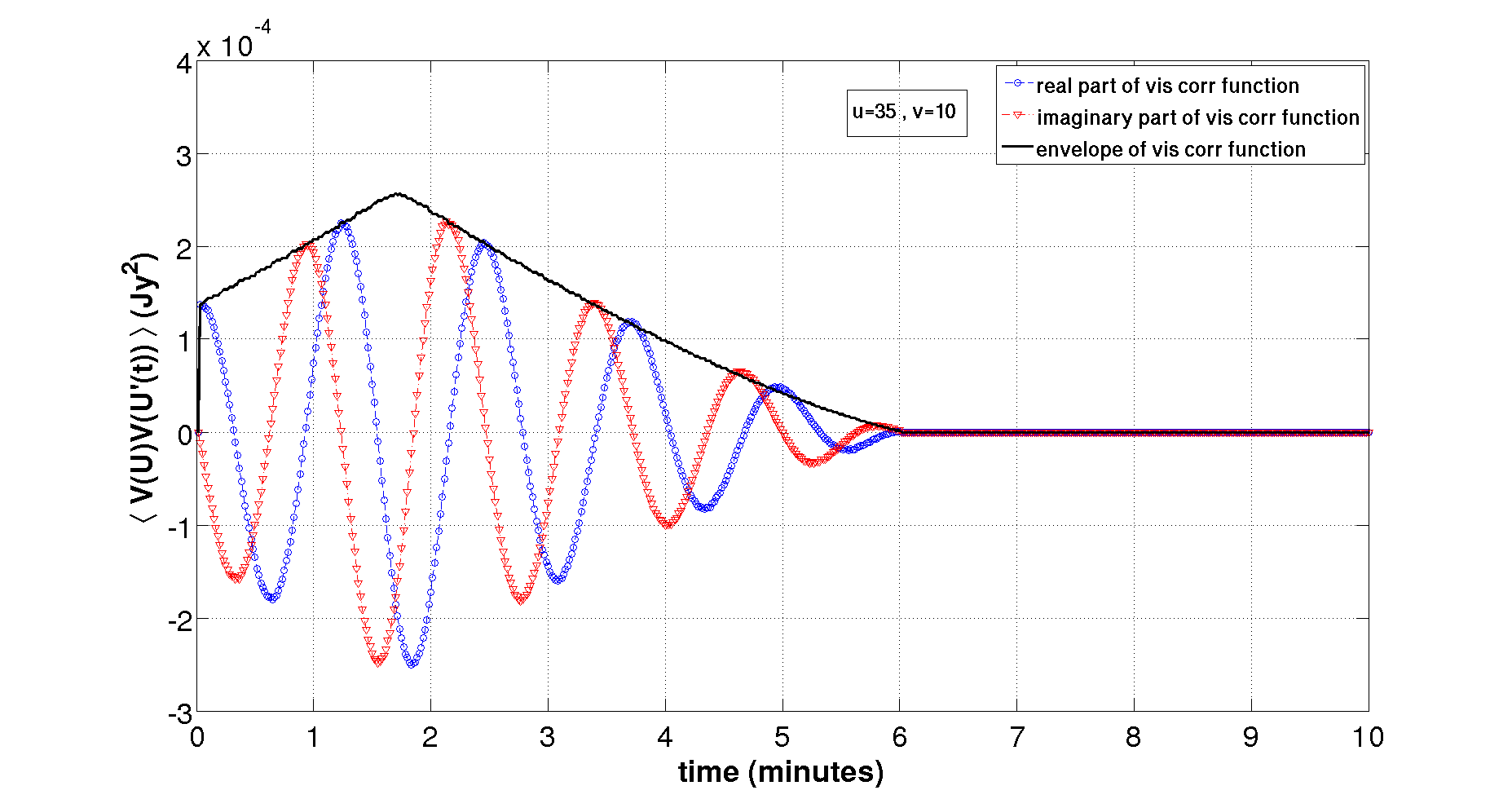}
\caption{\it Visibility correlation function as a function of time for visibilities with different baselines. The drift scan correspond to a zenith scan for 
a latitude of $30^\circ$}
\label{VV'}
\end{figure}

\begin{figure}[H]
\centering
\includegraphics[width=0.8\textwidth]{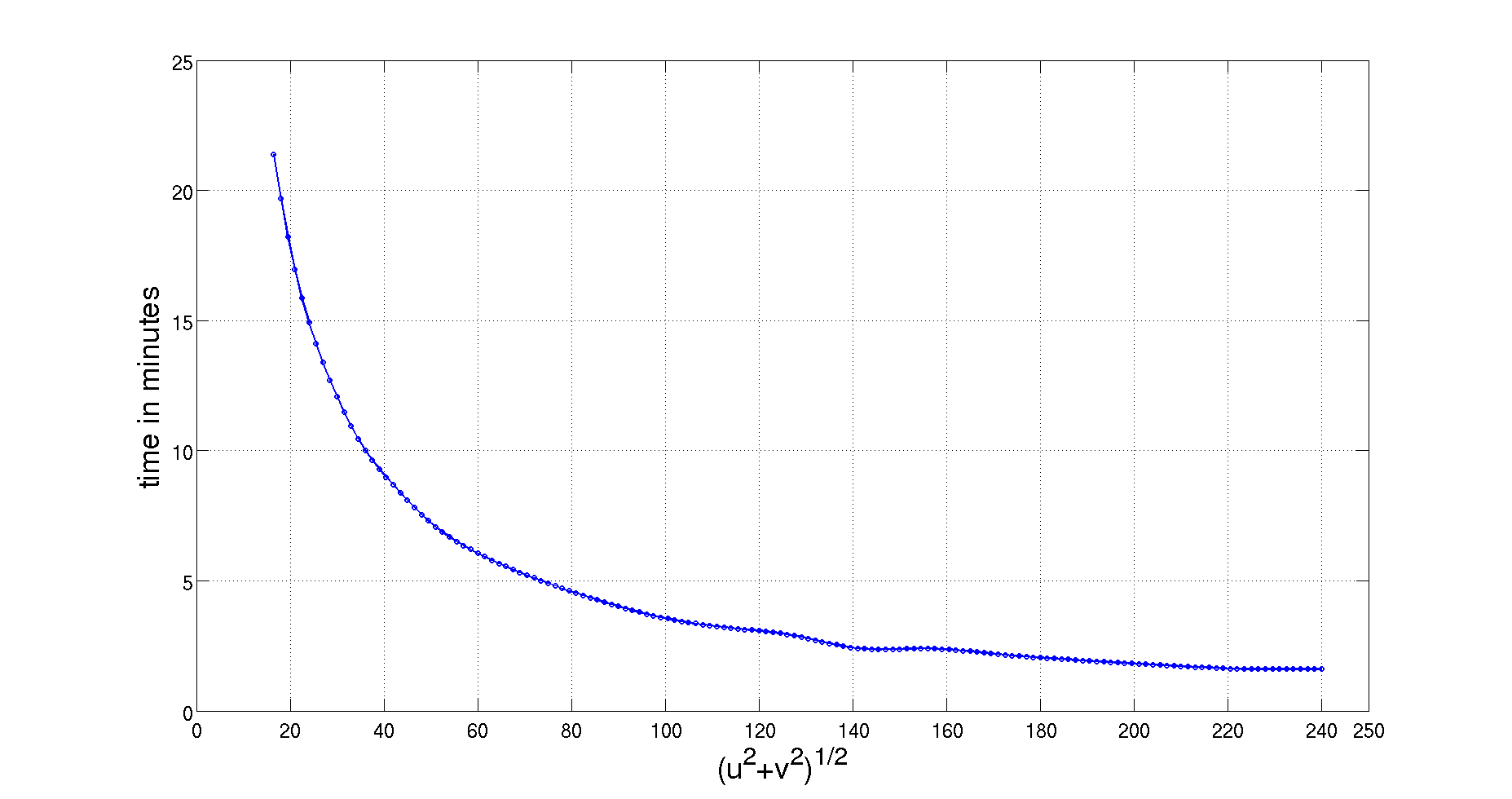}
\caption{\it  The Figure displays the time difference $\Delta$t at which
the visibility correlation  falls to half its value, as a function of 
baseline,  for an overhead scan
at the location of MWA ($\phi = -26.7$).}
\label{VV2}
\end{figure}

In Figure~\ref{VV2} we show the time scale over which the visibility
correlation  falls to  half its  value  for $\Delta t = 0$ (e.g. Figure~2);
this time scale is seen to fall as roughly the inverse of the baseline length,
in agreement with the discussion in the previous sub-section.

The re-correlation of baselines allows us to partially 
recover   the loss of signal due to decorrelation. However,
the set of baselines is fixed for a drift scan strategy and therefore 
the range of baselines that correlate at different times must be present 
in the initial set. For the MWA, we estimate that for a zenith scan at the 
latitude of the telescope there are nearly  120 such pairs which satisfy 
$\varepsilon \le 0.5$ and $a \le 0.3$ for $|{\bf U}| \simeq 20\hbox{--}70$.
 (as noted below, the total number of baselines in a zenith snapshot observation
for MWA is 2735 in the range $|{\bf U}| \simeq 20\hbox{--}230$.)
 These baselines will retain at least
half the signal and would  correlate within a correlation 
 time scale of less than  two hours.

\section{Error on visibility correlation}

The error on visibility  correlation is:
\begin{equation}
\sigma^2(U) = \langle V_\nu({\bf U}) V_\nu({\bf U}) V_{\nu'}({\bf U'},t) V_{\nu'}({\bf U'},t)  \rangle - \langle V_\nu({\bf U}) V_{\nu'}({\bf U'},t) \rangle^2
\end{equation}
Here $U \equiv |{\bf U}|$ and the averages   are taken over many 
different variables: the noise is uncorrelated for different frequencies, 
baselines, and times. However, the signal could be correlated in all the 
three domains. We average over all the pairs in the three domains 
and finally over  all the pairs for baselines in the  range $U$ and $U + {\Delta}U$ to compute an estimate for a wider bin $\Delta U$.  
The measured visibilities
and their  correlations receive contributions from detector noise, the HI signal, and the foregrounds. When only visibilities at two times (or frequencies/baselines) are correlated, 
as we assume here, the $\langle V V \rangle$ doesn't receive any contribution
from detector noise and therefore constitutes an unbiased estimator of the 
signal. In this case, only the first term in the equation above contributes
to the error estimate; denoting the sky noise as $N_\nu$, we get:
\begin{equation}
\sigma^2(U) = {1 \over N_{\rm tot}} \langle N_\nu({\bf U}) N^*_\nu({\bf U}) \rangle^2
\end{equation}
Here $N_{\rm tot}$ are all the   baseline pairs in the 
range $U$ and $U + {\Delta}U$ in the three-dimensional cube and the time domain. 

The average noise autocorrelation for 
 each independent correlation of visibilities  is:
\begin{equation}
\langle N_\nu({\bf U})N^*_\nu({\bf U})\rangle=\left[\frac{T_{\rm sys}}{K\sqrt{\Delta\nu\Delta t}}\right]^2 \label{NN}
\end{equation}
where $T_{\rm sys}$ is the system temperature, $\Delta\nu$ is the channel width, K is the antenna gain and $\Delta t$ is the integration time. Here 
$\Delta t$ and  $\Delta \nu$ could be arbitrarily small; in particular 
we require the bandwidth and integration time to be much smaller than
the frequency and time coherence of the signal (Figure~1--3 and~6).
 $N_{\rm tot}$ is  determined from the correlation times scale
in time and frequency domains and its computation is discussed below.
 We  cross-correlate
 all visibility pairs for a given time difference and frequency difference 
for the equatorial scan case  where we assume ${\bf U} = {\bf U'}$;  we also 
include 
the impact  of re-correlating baselines (section~3.1)  in  
the overhead scan case (Figure~2).  

\begin{figure}[h]
\centering
\includegraphics[width=0.6\textwidth]{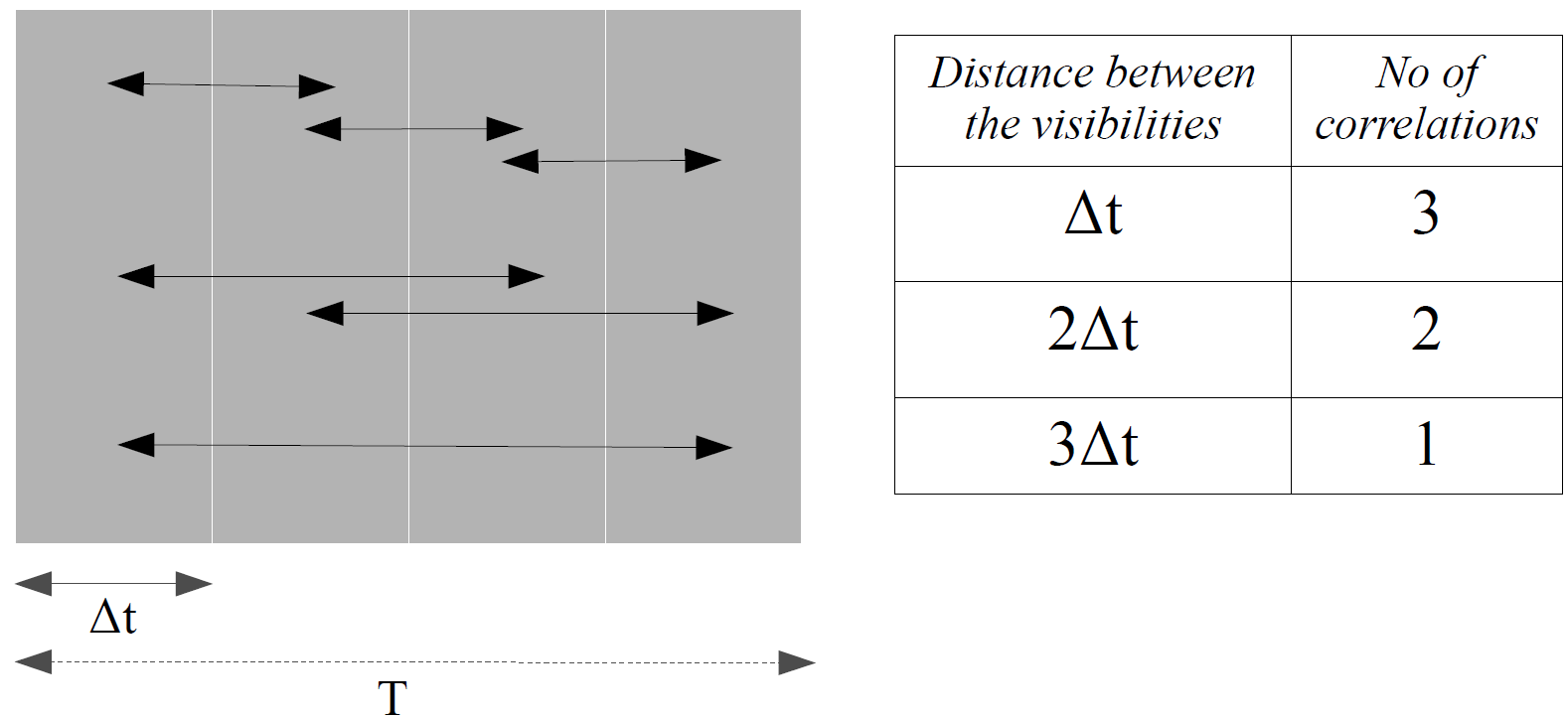}
\caption{\it Illustration showing the number of possible correlations for total observing time T and integration time 
$\Delta t$ with $T/\Delta t=4$, or  four visibility measurements. The number of correlations between visibilities with time difference $\Delta t$
is four, for time difference $2\Delta t$ the number is three and so on.}
\label{cartoon1}
\end{figure}
\vspace{10 pt}

For a  given total observing time T and integration time $\Delta t$, there exists $n=T/\Delta t$ visibility measurements. Among these, the 
number of possible independent correlations between visibilities $i\Delta t$ time apart (where i=1,2,....,n) is $(n-i)$ as explained in
figure~\ref{cartoon1}.
Thus, average noise correlation for a given baseline vector {\bf U} with visibilities separated by times $i\Delta t$ is:  
\begin{equation}
\sigma_i({\bf U})=\langle N_\nu({\bf U})N^*_\nu({\bf U}(i\Delta t))\rangle=\frac{1}{(n-i)}
\left[\frac{T_{sys}}{K\sqrt{\Delta\nu\Delta t}}\right]^2 \label{sigmai},
\end{equation}
for any frequency channel.

Figure~\ref{vis_t_mwa}  shows 
that the signal $\sqrt{\langle V_\nu({\bf U})V^*_{\nu'}
({\bf U'},t)\rangle}$ decorrelates with increasing time difference between the visibilities. This means that not all pairs contribute equally to the 
signal-to-noise of the measurement.  To obtain an estimator that 
gives suitable weight to all the pairs we define:
\begin{equation}
w_i({\bf U})=\frac{\langle V_\nu({\bf U})V^*_{\nu}({\bf U}(t=0))\rangle}{\langle V_\nu({\bf U})V^*_{\nu}({\bf U}(t=i\Delta t))\rangle} 
\end{equation}
This allows us to write the following optimal 
estimator for computing the noise on the  visibility measurement:
\begin{equation}
 \frac{1}{(\sigma_U^2)^2}=\displaystyle\sum_{i=1}^{n}\frac{1}{(\sigma_i^2 w_i)^2}
 \label{sigmau}
\end{equation}
We  neglect the effect of partial coherence of baselines 
at the initial time; this assumption slightly underestimates the 
sensitivity and  is further discussed 
 in the next subsection.  We do not include sampling variance in our error
estimates.

The foregoing discussion is valid for visibility measurements
for a given frequency. The HI signal is correlated across frequency 
space (Figure~\ref{freqcorr}).   
The figure shows  the behaviour of the HI signal as a function of $|\nu'-\nu|$
for different baselines. The right panel of the 
figure displays  the frequency difference  at which the signal falls 
to half of the of the maximum ($|\nu'-\nu|=0$) for different times. We treat the correlation across
 the frequency space using the same method described above  for 
 the   time correlation.  
\begin{figure}[h]
\centering
\includegraphics[width=1.0\textwidth]{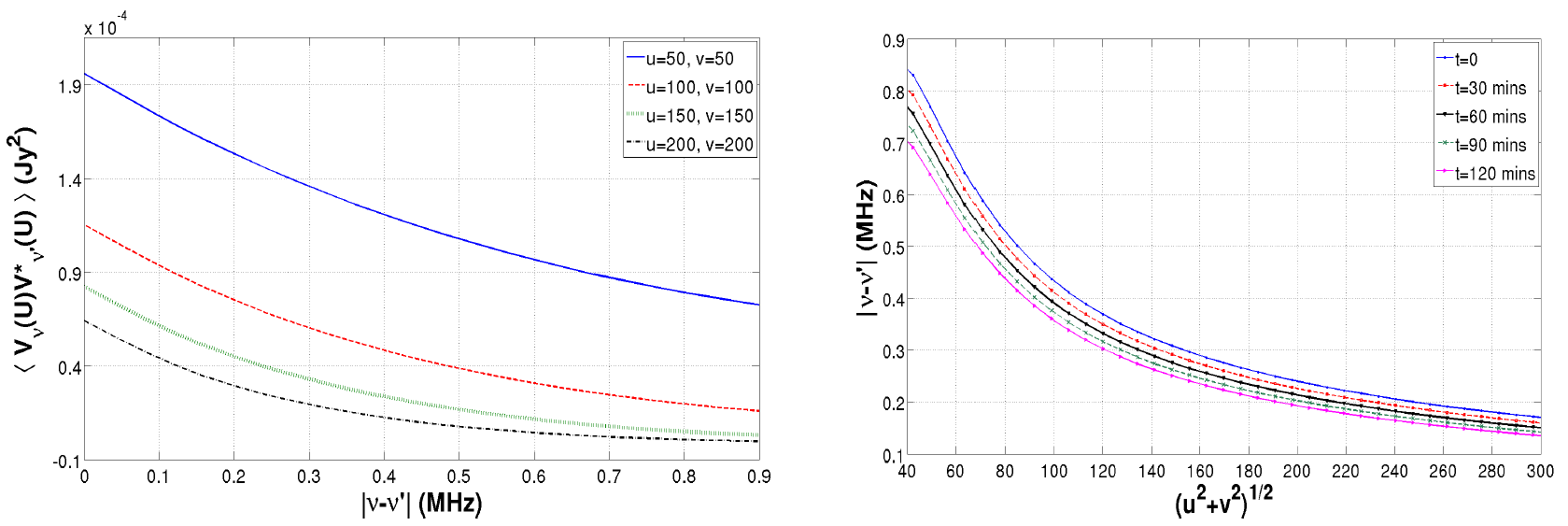}
\caption{\it The left figure shows the decorrelation of visibility correlation
of the HI signal as a  function of  frequency separation (see the 
caption of Figure~1 and the discussion in section~3 for details). The right panel  denotes the bandwidth for a given {\bf U} at which the signal drops to half of its maximum at different times}
\label{freqcorr}
\end{figure}

In Figure~\ref{noise}, we show the expected noise on the visibility correlation for many different cases. For all the cases, we assume the following parameters
for the MWA: observing frequency $\nu=129$ MHz, system temperature
 $T_{\rm sys}=  440$~K, and the effective   area of each tile $A_{\rm eff} = 16 m^2$.
\vspace{10 pt}

{\bf Case-I:} We consider a continuous  equatorial drift scan of 
a duration of  2 and~4 hours. One way  to repeat the 
scan for the same phase center is to shift the phase center to the same 
position after the end of the scan; this results in the change of UV coverage. 
We consider the simpler case when
the UV coverage and the phase center remain the same for subsequent 
scans. This corresponds to  the same region of sky being  observed   
 on different  days. In Figure~\ref{noise} we show the results
for 900~hours of integration in this mode. 

As the  signal strength is greater 
for shorter baselines (figure~\ref{vis_t_mwa}),  we consider only baselines  in the 
range ${\bf U}=20\hbox{--}230$. We take bins of size ${\bf U}\simeq 10$ and
 show the noise
correlation for this range of baselines  in Figure~\ref{noise}. 
 MWA has 2735 baselines  in this range for a snap-shot observation. 
Using the information,  the
RMS noise for this mode is $\sigma \simeq 16$~(mJy)$^2$ and $\sigma \simeq 21$~(mJy)$^2$ for 2~and~4 hours scan, respectively. We note that since the visibility
 correlation function
 drops significantly after roughly  1 hour (Figure~\ref{eq50}), the noise is 
expected to  increase for longer drift scans. 
\vspace{10 pt}

{\bf Case-II:} Here we consider an overhead drift scan at the  location
of MWA. The correlation time scale is shorter for such scans as compared to the 
equatorial scan (Figure~\ref{30deg50}). In Figure~\ref{VV2} we show 
the time scale over which the correlation falls by half as 
a function of the baseline length. 

As noted above many baselines get re-correlated
as the time progresses (Figure~\ref{VV'}). Over 5--10\% of all
 the baselines in the range $|{\bf U}| = 20\hbox{--}100$ get re-correlated 
with $\epsilon \le 0.5$ in less than two hours. We include these baselines in
the noise computation. As compared to Case~I, the noise is higher in this 
case as the correlation time is shorter. 
\vspace{10 pt}

{\bf Case-III}: For comparison with the drift scan cases, we also 
compute the error in the visibility correlation for the tracking case.
We consider two cases: 2~and 6~hours continuous  tracking of 
 a region across the zenith 
($\pm 1$h and $\pm 3$h) at the location of MWA ($\phi = -26.7$). The 
results are shown Figure~9. We discuss the results in detail in the next 
subsection.

\begin{figure}[H]
\centering
\includegraphics[width=1.0\textwidth]{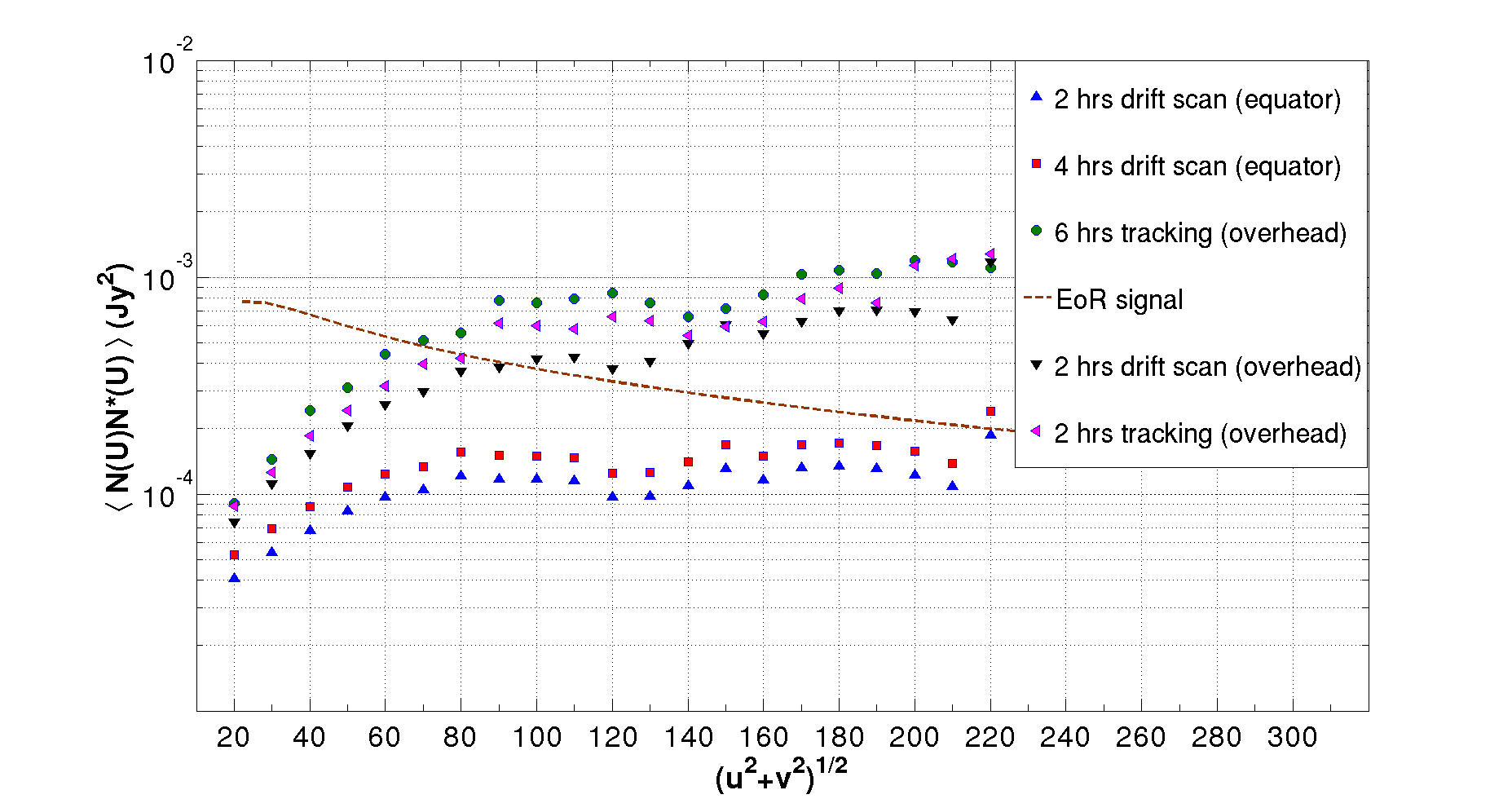}
\caption{\it Error  on visibility  correlation
 as a function of baseline length: blue (triangle) and  red (square) points
 refer to 2 and  4~hours equatorial drift scans, 
respectively. Black
(inverted triangle)   points refer to 2~hours  zenith drift scan
at the location of MWA. The green  (circle) and pink (rotated 
triangle) points  show the expected error for 2~and 6~hour tracking runs (for $\pm 3$ and $\pm 1$ hour
 continuous overhead tracking at MWA location). In all the cases the total integration is 
 900 hours. The EoR signal is designated by the dashed brown line.}
\label{noise}
\end{figure}

\subsection{Drift vs tracking mode}
In any interferometric experiment to determine the EoR signal,
 the RMS noise on the visibility 
correlation  is bounded by:
\begin{eqnarray}
\sigma_{\rm min} & = & \left ({1\over N_b} \right )^{1/2} \left ( {T_{\rm sys}  K \over \sqrt{\Delta \nu T}} \right )^2 \\
\sigma_{\rm max} & = & \left ({\Delta t \over N_b T}\right )^{1/2} \left ( {T_{\rm sys} K \over \sqrt{\Delta \nu  \Delta t}}\right )^2. 
\end{eqnarray}
Here $T$ is the total time of integration and $\Delta t$ is the integration
time for a single visibility measurement. For the sake of the discussion, 
$\Delta \nu$, the channel width is assumed to be fixed. $N_b = n(n-1)/2$ is 
the total number of baselines for any measurement with n antenna elements. $\sigma_{\rm min}$ gives
the RMS noise if all the visibilities are coherently added and $\sigma_{\rm max}$ corresponds to the case when the visibility  correlations 
are incoherently added.  For the 128-tile MWA, the RMS lies between these two extremes
for both the tracking mode and drift scans. As noted above, we neglect 
partially coherent baselines for computing the sensitivity for  drift scans;
this assumption is consistent with Eq~(23).

 The process of decoherence occurs differently for the tracking 
and the drift scan mode. For drift scans, it is decorrelation of the EoR
signal 
at different times, as described in detail in the previous sections. In 
the tracking case, the process of tracking a given region rotates
 the visibility
vector ${\bf U}$; the correlation between visibility measurements at
different values of  ${\bf U}$ decreases; from Eq.~(\ref{crosscor}), we can show
that the decorrelation scale $\Delta {\bf U} \simeq \theta_0^{-1} \simeq 0.5$
has very weak dependence on the value of ${\bf U}$. For our computation
we take the pixel size: $\{\Delta U, \Delta V\} = \{0.5, 0.5\}$.  The frequency
decorrelation for the tracking case is taken from figure~\ref{freqcorr}. 

The results for  two and six hour tracking runs  (zenith at the location of 
MWA)  are shown for 900 hours of integration in figure~\ref{noise}. We note here
that  we do not present the results for equatorial tracking run, as the 
sensitivity in this case shows only a marginal improvement over the zenith
tracking runs  shown in Figure~9. 
 As the figure shows, the
 drift scan generally gives lower noise on  the visibility  correlation 
for up to  4~hours  of drift scans.

This result can be understood as follows. As an extreme case, one could 
drift for a very short duration each day, such that there is no decorrelation
and continue similar observations on the same field  such that
all the visibility measurements are coherently added.
In this case, the RMS for the drift case would approach $\sigma_{\rm min}$
which is not possible to achieve in the tracking case because the process
of tracking would always decorrelate the signal. The relevant question is: 
what is the time scale 
for drift scans such that this advantage of lower noise is not
lost. We show that even for four-hour drift scans this advantage holds. 
In the drift scan case, the decorrelation time scale is $\simeq 1$hour. In the 
tracking cases, different baseline decorrelate in the process of tracking 
a region of the sky but some baselines revisit the same pixel in this process.
For instance, for a six hour tracking run shown in Figure~9,  the average integration time of a pixel in the range: $|{\bf U}| = 20\hbox{--}30$ is roughly 15 minutes with the total number of
uncorrelated  pixels  $\simeq 4700$.

It should be underlined that, apart from other 
assumptions delineated in the previous sub-section,  the lower noise
 in the drift scan is also based on the 
assumption that the system temperature doesn't change over the scan.  Also an
 additional disadvantage in the drift scan case is that there are
 smaller number 
of visibility measurements   available at any given time for imaging as 
compared to the tracking case where the UV coverage is better. 

\section{Statistical homogeneity of EoR signal and foreground extraction}
Unlike the tracking case, the drift scans explicitly exploit the statistical
homogeneity of the EoR signal: cross correlation of the signal at 
different times  only depends on the time difference. More precisely, the 
power spectrum of the EoR signal for any phase center is drawn from 
a random density field with the same average power spectrum. This assumption
may or may not hold for foregrounds. For instance, if faint point sources
are distributed homogeneously across the sky with the same flux distribution, 
they will also closely correspond to a statistically homogeneous field in two
dimensions. However, most other foregrounds, e.g. bright point sources 
or galactic foregrounds, will explicitly break the statistical homogeneity
of the sky and therefore would be potentially distinguishable from the 
EoR signal.
\vspace{10 pt}

We illustrate this concept with point source distribution on the sky.  For
a point source distribution with fluxes $\{F_i\}$, the visibility can 
be written as:
\begin{equation}
V({\bf U}) = \sum_j \exp(2\pi i{\bf U}.{\bf \theta_j}(t)) F_j A({\bf \theta_j}(t))
\end{equation}
Here ${\bf \theta_i}(t)$ correspond to the time varying 
 position of point sources on the sky with respect to the fixed phase center. 
$A({\bf \theta_i}(t))$ gives the primary beam in the same coordinate system. The visibility   correlation separated by time $\Delta t$ is: 
\begin{equation}
\langle V({\bf U}, t)V^*({\bf U}, t+\Delta t) \rangle  = \langle  \sum_k \sum_j \exp(2\pi i {\bf U}.({\bf \theta_j}-{\bf \theta_k})) F_j F_k 
 A({\bf \theta_j}(t)) A({\bf \theta_k}(t+\Delta t)) \rangle 
\end{equation}
Here the averaging process $\langle\cdots\rangle$ is over all the pairs
for a given $\Delta t$ during the drift scan. This averaging procedure 
 leads to substantially  different results for the EoR signal
and the foregrounds: the EoR signal is statistically homogeneous and therefore
any cross-correlation depends only on $\Delta t$. For 
each $\Delta t$ the EoR signal gives a realization of the density field
with a given fixed  power spectrum (Eq.~(\ref{crosscor})). However, the
 foregrounds 
might not share this property and might show explicit dependence not
just the time difference but the time period of the scan. This gives 
at least  two different methods of extracting foregrounds:
(a) correlation pairs of a given $\Delta t$  can be used to fit the  time variation 
expected of foregrounds. While the EoR signal will show fluctuations about 
a given mean, the foregrounds will show more secular time variation which
can potentially be subtracted, (b) direct comparison of the averaged 
correlation function should also reveal the difference between the two cases. 
We demonstrate the procedure with method (b) here.
\vspace{10 pt}

For MWA primary beam, we consider 
 three different source counts: 10, 30 and 50~sources. At the 
beginning of the drift scan the sources are randomly distributed within \textpm$15^\circ$ from the center of the primary beam with hour angle  
between -3 to +3 hours. The fluxes are drawn from uniform 
distribution with values  between 0~and 1~Jy.
The visibility correlation function for all these cases for  {\bf U}=(50,50) are shown in figure~\ref{pt_source_corr}.

\begin{figure}[H]
\centering
\includegraphics[width=0.9\textwidth]{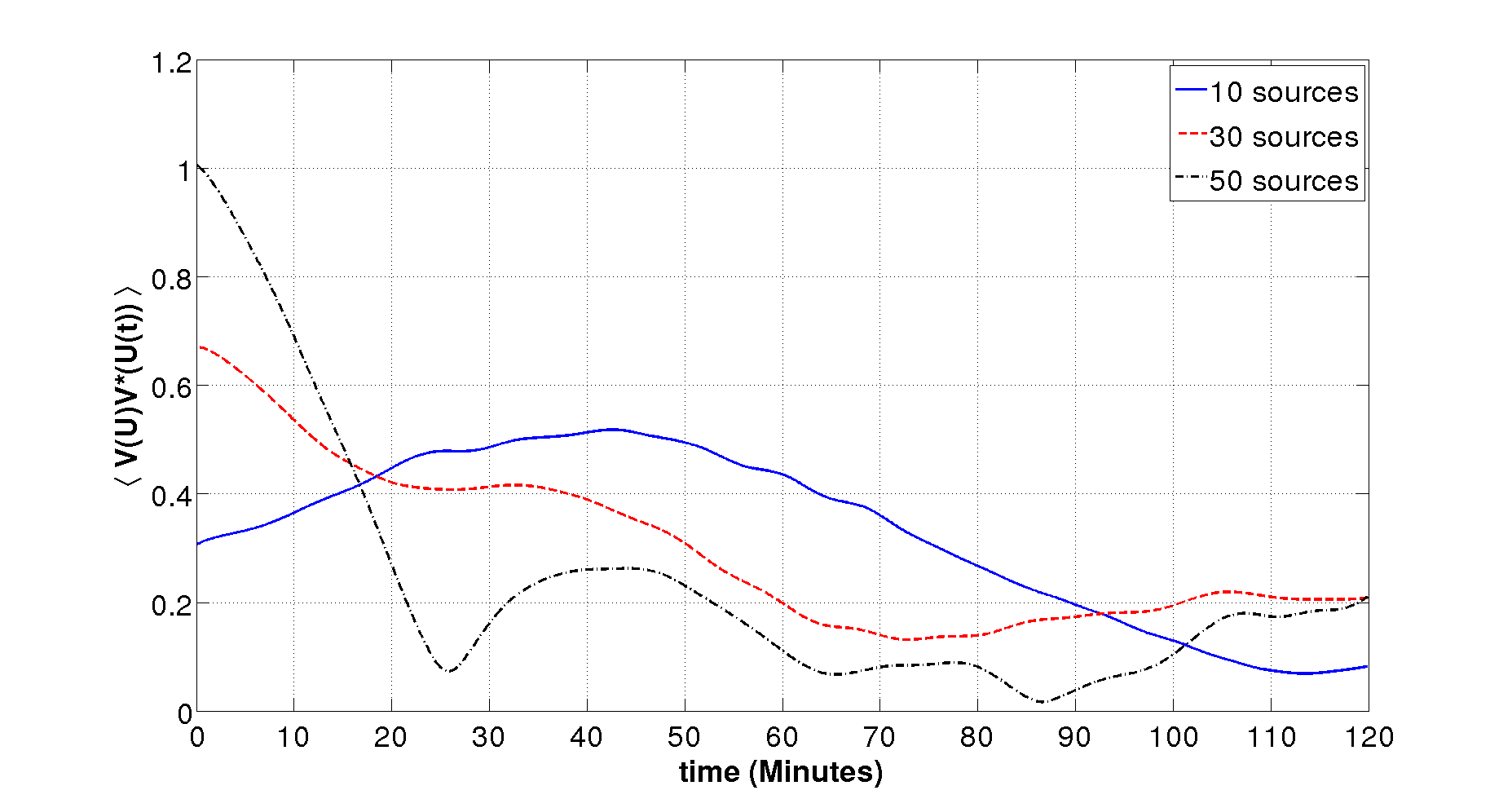}
\caption{\it Envelope of the Visibility correlation function (normalized arbitrarily) as a function of time difference for three different 
cases described in the text.}
\label{pt_source_corr}
\end{figure}
As predicted in the foregoing, figure \ref{pt_source_corr} shows that 
 the visibility correlation function for point sources 
is substantially different  as compared to the 
HI signal owing  to
 statistical inhomogeneity of  the point source distribution. 
This can be used to subtract the contribution of bright point sources from the measured visibility itself.

\section{Conclusions}
The main goal of this paper is to investigate the potential of  the drift
 scan technique in estimating  the EoR signal.
Drift scans introduce a new dimension to the issue: 
the correlation between visibilities in time domain.  Here we present
a formalism which uses this correlation to determine the EoR signal. 

The important results are as follows:
\begin{itemize}
 \item The visibilities measured at different times by the same pair of antennas
 in a drift scan are correlated for up to 1~hour for equatorial scans (Figure~\ref{eq50}).   
 The decorrelation time scale depends on the choice of phase center.
 It is maximum for an equatorial zenith drift or for  equatorial 
phase center.  For such scans, the decorrelation time scale is independent of 
the baseline length. For other scans  the 
decorrelation times scale is shorter and depends on the baseline vector
 (Figure~\ref{30deg50}--\ref{pole50}). However, a fraction of these baselines
correlate with other baselines at a different time (Figure~\ref{VV'}). 

\item We compute the  expected error  on the visibility  correlation
for drift scans
and  compare with the  expected noise in the tracking case
(Figure~\ref{noise}). Our results show that the noise is comparable in the 
two cases and the drift scan might lead to a superior signal-to-noise 
for equatorial scans.

 \item The drift scan technique also opens another avenue  for 
the extraction and subtraction of foregrounds: the EoR signal is
 statistically homogeneous while the foregrounds might not share this 
property. We investigate the potential of this possibility using a set of 
bright point sources (figure~\ref{pt_source_corr}). 
\end{itemize}

Our results suggest that drift scans might provide a viable, and potentially
superior, method for extracting the EoR signal. In this 
paper, we present mainly analytic results to make our case. In the future, 
we hope to return to this issue with numerical simulations and 
 direct application of our method to  the MWA  data. 
 
 \vspace{20 pt}
 {\bf Acknowledgments:} We thank Rajaram Nityananda, Ron Ekers, Sanjay Bhatnagar, Urvashi Rau, 
 Nithyanandan Thyagarajan, Subhash Karbelkar for the useful comments and discussions. We thank the referee for penetrating comments which helped us to  improve
 the paper. 
 
 This scientific work makes use of the Murchison Radio-astronomy Observatory, operated by CSIRO.
 We acknowledge the Wajarri Yamatji people as the traditional owners of the Observatory site. 
 Support for the MWA comes from the U.S. National Science Foundation (grants AST-0457585, PHY-0835713, CAREER-0847753, and AST-0908884),
 the Australian Research Council (LIEF grants LE0775621 and LE0882938), the U.S. Air Force Office of Scientic Research (grant FA9550-0510247),
 and the Centre for All-sky Astrophysics (an Australian Research Council Centre of Excellence funded by grant CE110001020). 
 Support is also provided by the Smithsonian Astrophysical Observatory, the MIT School of Science, the Raman Research Institute,
 the Australian National University, and the Victoria University of Wellington (via grant MED-E1799 from the New Zealand Ministry of Economic
 Development and an IBM Shared University Research Grant). The Australian Federal government provides additional support via the Commonwealth
 Scientific and Industrial Research Organisation (CSIRO), National Collaborative Research Infrastructure Strategy, Education Investment Fund,
 and the Australia India Strategic Research Fund, and Astronomy Australia Limited, under contract to Curtin University.
 We acknowledge the iVEC Petabyte Data Store, the Initiative in Innovative Computing and the CUDA Center for Excellence sponsored by
 NVIDIA at Harvard University, and the International Centre for Radio Astronomy Research (ICRAR), a Joint Venture of Curtin University and
 The University of Western Australia, funded by the Western Australian State government. 

\newpage
\section{Appendix}
\begin{appendix}
\section{Coordinate system for drift scans} \label{AppendixA}
The position vector in the sky $\vec\theta$ can be expressed in terms of two
 direction cosines l and m. These direction cosines are defined
with respect to a local coordinate system when phase center is at zenith  as explained in figure~\ref{lmn} (e.g. \cite{chris}):
\begin{eqnarray}
 l & = &\cos\delta \sin H \nonumber \\
 m & = &\cos\delta\cos H\sin\phi-\sin\delta\cos\phi \nonumber \\
 n & = &\cos\delta\cos H\cos\phi+\sin\delta\sin\phi \label{lmneqn}
\end{eqnarray}
Here $\delta$ and H are the declination and hour angle of any source; and $\phi$ is the latitude of the place of observation.
\begin{figure}[H]
\centering
\includegraphics[width=0.5\textwidth]{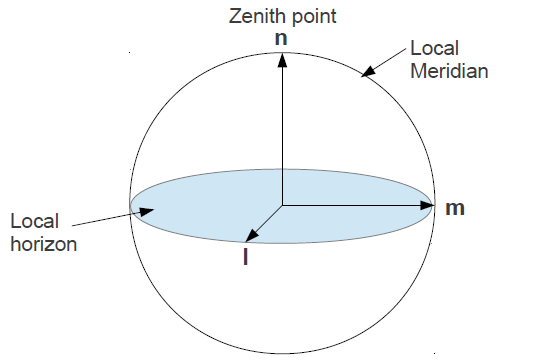}
\caption{\it l,m,n coordinates defined for a phase center at zenith}
\label{lmn}
\end{figure}

Using this coordinate system the second integral in visibility expression
Eq.~(\ref{viscorrt}) takes the form:
\begin{equation}
\int dl dm A(l,m)\exp\left[-2\pi i\left\{\left(u-\frac{k_{\perp 1}r_\nu}{2\pi}\right)l+\left(v-\frac{k_{\perp 2}r_\nu}{2\pi}\right)m\right\}
\right]\exp\left[-ir_\nu\left(k_{\perp 1}\Delta l +k_{\perp 2}\Delta m\right)\right] 
\end{equation}
Here $\Delta l$ and $\Delta m$ are the change in $l$ and $m$ with time or
 hour angle as in sky drift only hour angle changes with time for a fixed
declination. 
$k_{\perp 1}$ and $k_{\perp 2}$ are the two components of ${\bf k}_\perp$ along $l$ and $m$ on the sky plane. Using Eq.~(\ref{lmneqn}) 
and the condition $l^2+m^2+n^2=1$
we can expand in the first order to compute the changes in 
relevant quantities: 
\begin{equation}
\Delta l= (m\sin\phi+n\cos\phi)dH; \hspace{15 pt} \Delta m=-(l\sin\phi) dH.
\label{rottran}
\end{equation}
Here dH is the change in hour angle in time interval $t$. 
We can further simplify the expression by using $n \simeq 1$. 
The two approximation used above are: $1/2(l^2 + m^2) \ll 1$ and 
$dh \ll 1$. Both these approximations are valid for 
the MWA primary beam (Eq.~(\ref{pribeam})) and for a few hours of correlation
time.   Thus the second integral (Eq.~\ref{lmneqn}) becomes:
\begin{eqnarray}
 \exp\left(-ir_\nu k_{\perp 1}\cos\phi dH\right)\int dldm A(l,m)\exp\left[-2\pi i\left\{\left(u-\frac{r_\nu}{2\pi}(k_{\perp 1}+k_{\perp 2}\sin\phi
 dH)\right)l\right.\right. \nonumber \\
 \left.\left.+\left(v-\frac{r_\nu}{2\pi}(k_{\perp 2}-k_{\perp 1}\sin\phi dH)\right)m \right\}\right] \nonumber
\end{eqnarray}
It can be expressed in terms of the Fourier transform of the primary beam:
\begin{equation}
 \exp\left(-ir_\nu k_{\perp 1}\cos\phi dH\right) a\left[\left(u-\frac{r_\nu}{2\pi}(k_{\perp 1}+k_{\perp 2}\sin\phi dH)\right)
 ,\left(v-\frac{r_\nu}{2\pi}(k_{\perp 2}-k_{\perp 1}\sin\phi dH)\right)\right]
\end{equation}

With this the visibility measured at a later time $t$ becomes:
\begin{eqnarray}
 V_\nu({\bf U},t)&=&\bar{I}_\nu\int\frac{d^3k}{(2\pi)^3}\Delta_{HI}({\bf k})e^{ir_\nu k_\parallel}\exp\left(-ir_\nu k_{\perp 1}
 \cos\phi dH\right) \nonumber \\
 &&\:a\left[\left(u-\frac{r_\nu}{2\pi}(k_{\perp 1}+k_{\perp 2}\sin\phi dH)\right)
 ,\left(v-\frac{r_\nu}{2\pi}(k_{\perp 2}-k_{\perp 1}\sin\phi dH)\right)\right]
\label{vislattime}
 \end{eqnarray}
 Correlating this with the visibility measured at t=0 (equation \ref{viscorr}) gives:
 \begin{eqnarray}
  \langle V_\nu({\bf U})V^*_{\nu'}({\bf U'},t)\rangle &=& \bar{I_\nu}^2\int \frac{d^3k}{(2\pi)^3}P_{HI}(k)e^{ik_\parallel \Delta r_\nu}
  \exp\left(-ir_{\nu'} k_{\perp 1}\cos\phi dH\right) \nonumber \\
  &&\:a\left[\left(u-\frac{r_\nu k_{\perp 1}}{2\pi}\right),\left(v-\frac{r_\nu k_{\perp 2}}{2\pi}\right)\right] \nonumber \\
  &&\:a\left[\left(u'-\frac{r_\nu}{2\pi}(k_{\perp 1}+k_{\perp 2}\sin\phi dH)\right)
 ,\left(v'-\frac{r_\nu}{2\pi}(k_{\perp 2}-k_{\perp 1}\sin\phi dH)\right)\right] \nonumber \\ \label{vvlmn}
 \end{eqnarray}
Eq.~(\ref{vvlmn}) and the discussion in this section allows us to
interpret Figure~1--3. If $\phi = 0$, or the observatory is located at the 
equator, then the trajectory of sources around the phase center in a drift scan
is pure translation; for any non-zero $\phi$ the motion 
is a combination of rotation and translation  (Eq.~(\ref{rottran}). For 
pure translation, one obtains Figure~1, or the decorrelation time scale
is determined solely by the extent of the primary beam. The decorrelation 
time scale is shorter for any non-zero $\phi$ (Figure~2, 3, and~6) 
 and depends on the baseline, as already noted in section~3. 

MWA is not located at the equator but we show below that, even for an
observatory not located at the equator,  
 if the phase center is shifted to an  equatorial position 
 one can remove the rotation  of sources 
in the  coordinate system constructed for the new phase center. For 
simplicity we construct a coordinate system around  the local meridian but our
conclusions remain valid for any phase center along the equator. 
 
\vspace{10 pt}
 For a phase center that  lies on the local meridian  with angular separation $\theta$ from
the zenith at the observatory (Figure~\ref{lmn_prime}), the  new set of coordinate system is obtained by a single rotation $\theta$ of the m and n axes about l as shown in the 
figure~\ref{lmn_prime}.
\begin{figure}[H]
\centering
\includegraphics[width=0.5\textwidth]{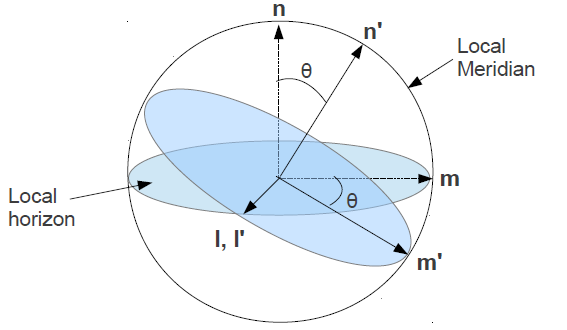}
\caption{\it Illustration of new lmn coordinate system}
\label{lmn_prime}
\end{figure}
Thus the new  coordinates can be expressed as:
\begin{equation}
 \begin{pmatrix}
  l' \\ m' \\ n'
 \end{pmatrix}=\begin{pmatrix} 1 & 0 & 0 \\ 0 & \cos\theta & \sin\theta \\ 0 & -\sin\theta & \cos\theta \end{pmatrix}
 \begin{pmatrix} l \\ m \\ n \end{pmatrix} \label{lmnmat}
\end{equation}

Substituting l,m,n values from equation (\ref{lmneqn}) we get:
\begin{eqnarray}
l'&=&\cos\delta\sin H \nonumber \\
m'&=&\cos\delta\cos H\sin(\theta+\phi)-\sin\delta\cos(\theta+\phi) \nonumber \\ 
n'&=&\cos\delta\cos H\cos(\theta+\phi)+\sin\delta\sin(\theta+\phi) \label{lmneqn_prime}
\end{eqnarray}

We illustrate the difference between the two 
coordinate systems with a set of point sources 
 with given initial positions (hour angle and declination) and compute  
source trajectories  in both lmn and l'm'n' coordinates. The unprimed
coordinates are for a zenith scan at the location of the observatory. 
The primed coordinates are for a phase center which is at equatorial position
at  the meridian. In this case,  for on observer situated at latitude $\phi$, the angle of rotation  $\theta=-\phi$. For instance for an observer at 
latitude $\phi=30^\circ$N, rotation angle is $\theta=-30^\circ$. 

Ten sources are chosen randomly within declination \textpm$10^\circ$ of the center of the primary beam and all with initial hour angle -2h.
The sources are allowed to drift past the primary beam for a total drift duration of 4 hours. The trajectories are shown 
figures~\ref{abc} and  \ref{def}. A contour plot of the primary beam is also included in each figure.

\begin{figure}[H]
\centering
\includegraphics[width=0.7\textwidth]{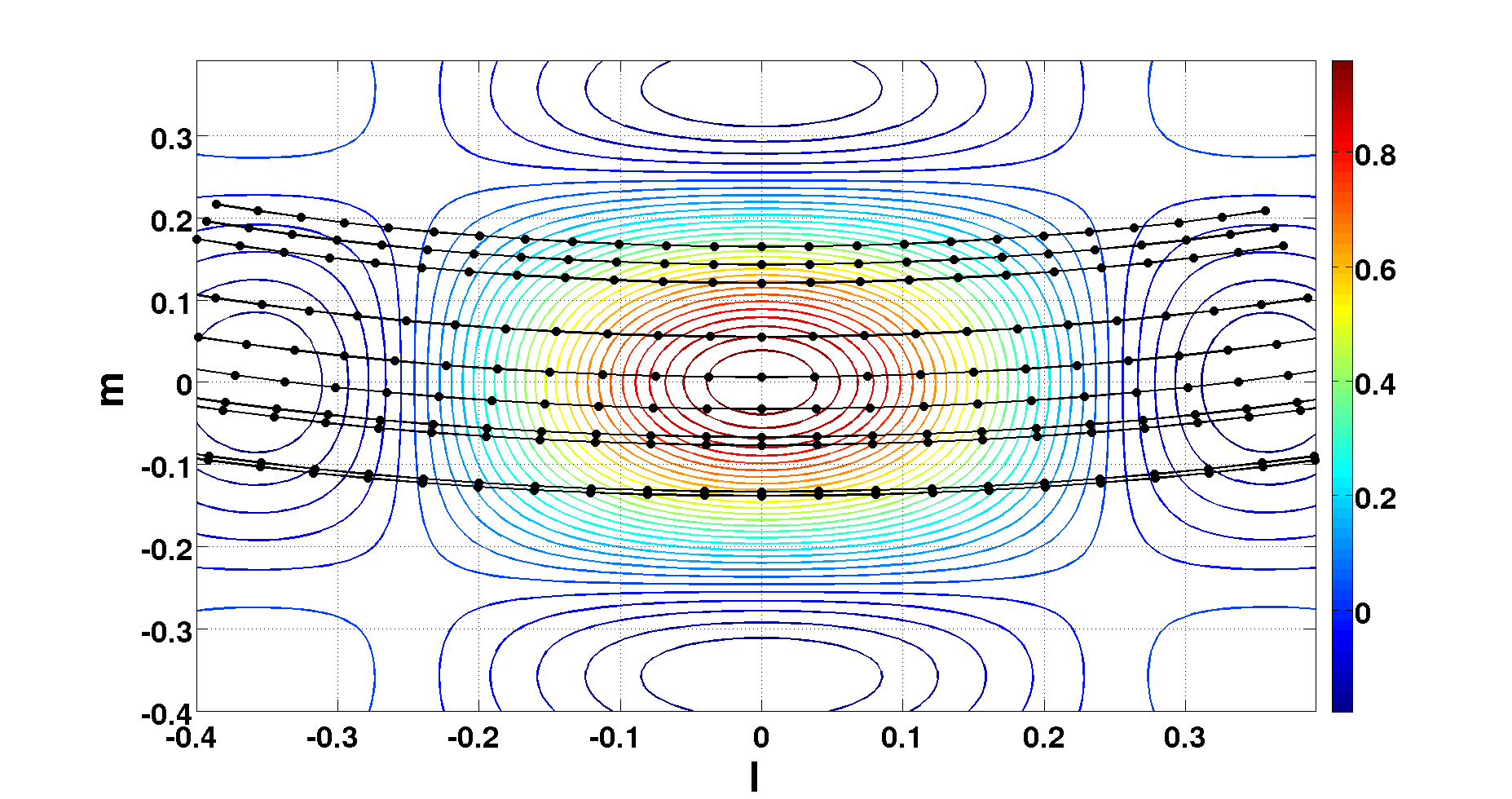}
\caption{\it source trajectories in lmn coordinate system (phase center at zenith) for an observer at latitude $-30^\circ$}
\label{abc}
\end{figure}

\begin{figure}[H]
\centering
\includegraphics[width=0.7\textwidth]{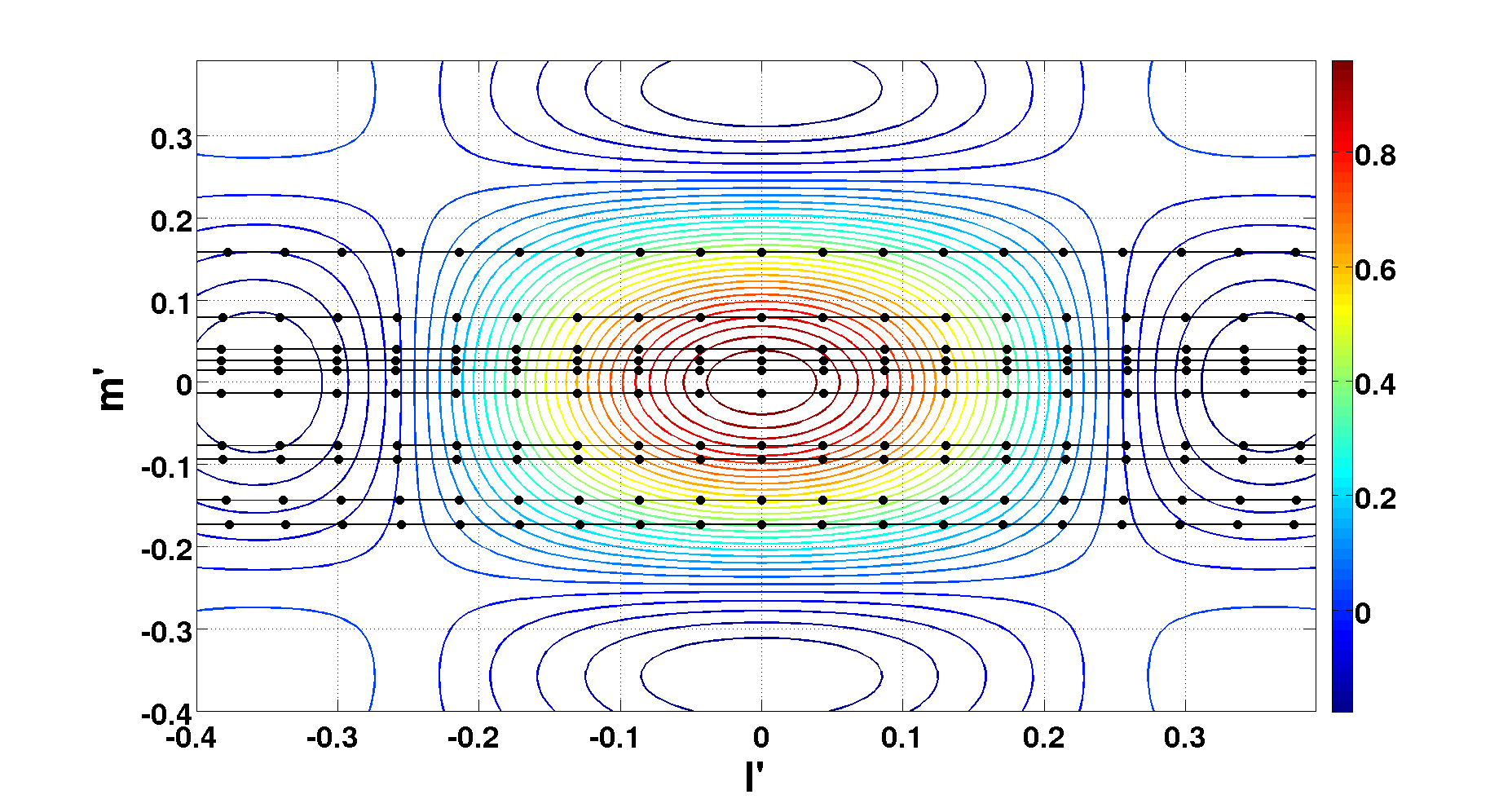}
\caption{\it source trajectories in l'm'n' coordinate system (phase center shifted to equator) for the  same observer  as in Figure~\ref{abc}}
\label{def}
\end{figure}

Using the new primed coordinate system instead of the previous one with phase center at zenith at the observatory,  one obtains the expression for the 
visibility 
correlation function as:
\begin{eqnarray}
  \langle V_\nu({\bf U})V^*_{\nu'}({\bf U'}, t)\rangle &=& \bar{I_\nu}^2\int \frac{d^3k}{(2\pi)^3}P_{HI}(k)e^{ik_\parallel \Delta r_\nu}
  \exp\left(-ir_{\nu'} k_{\perp 1}\cos(\theta+\phi) dH\right) \nonumber \\
  &&\:a\left[\left(u-\frac{r_\nu k_{\perp 1}}{2\pi}\right),\left(v-\frac{r_\nu k_{\perp 2}}{2\pi}\right)\right] \nonumber \\
  &&\:a\left[\left(u'-\frac{r_\nu}{2\pi}(k_{\perp 1}+k_{\perp 2}\sin(\theta+\phi) dH)\right)\right., \nonumber \\
  &&\:\left.\left(v'-\frac{r_\nu}{2\pi}(k_{\perp 2}-k_{\perp 1}\sin(\theta+\phi) dH)\right)\right] \nonumber \\ \label{vvlmn_prime}
 \end{eqnarray}
Eqs~(\ref{vvlmn_prime}) and~(\ref{vvlmn}) are the main results of the paper.

\section{w-term and other assumptions}

We have neglected the  w-term in our formalism. In this section, 
we attempt to assess the possible impact of  this 
term. The inclusion of w-term changes Eq~(\ref{visibility}) to: 
\begin{equation}
 V_\nu({\bf U})=\int A(\vec{\theta}) I_\nu (\vec{\theta})e^{-i2\pi (ul+vm+w(1-n) )} d\Omega\label{visibility1}
\end{equation}
Here $n = (1-l^2-m^2)^{1/2}$. The solid angle $d\Omega = dl dm/(1-n)$. 
For MWA primary beam we can use the flat sky approximation $1/2(l^2+m^2) \ll 1$
(Figure~13 and~14). As noted above  this approximation might break down 
when regions close to horizon are tracked. However, it remains a good approximation for zenith drift scan. We also make the simplifying assumption that
the primary beam is a Gaussian: $A(l,m) = \exp(-(l^2+m^2)/\theta_0^2)$; this 
allows us to make analytic estimates. 

From Eq.~(\ref{vvlmn_prime}), including the $w$ term,  the visibility at any time $t$ can be written as (we assume $\theta =0$, or a zenith scan):
\begin{eqnarray}
 V_\nu(u,v,w;t)&=&\bar{I}_\nu\int\frac{d^3k}{(2\pi)^3}\Delta_{HI}({\bf k})e^{ir_\nu k_\parallel}\exp\left(-ir_\nu k_{\perp 1} \cos\phi dH\right) \nonumber \\
 &\times& \int dldm A(l,m)\exp\left[-2\pi i\left\{\left(u-\frac{r_\nu}{2\pi}(k_{\perp 1}+k_{\perp 2}\sin\phi dH)\right)l \right.\right.\nonumber \\
& + &\left.\left.\left(v-\frac{r_\nu}{2\pi}(k_{\perp 2}-k_{\perp 1}\sin\phi dH)\right)m -{1\over2}w(l^2+m^2) \right\}\right] \nonumber
\label{vislattime1}
 \end{eqnarray}
For a Gaussian primary beam, the integral over angles can be computed 
analytically by extending the integration limits from $-\infty$ to $\infty$
which is permissible as the primary beam has a narrow support. This gives us:
\begin{eqnarray}
 V_\nu(u,v,w;t)&=&\bar{I}_\nu\int\frac{d^3k}{(2\pi)^3}\Delta_{HI}({\bf k})e^{ir_\nu k_\parallel}\exp\left(-ir_\nu k_{\perp 1} \cos\phi dH\right) \nonumber \\
 &\times& \left ({\pi \over q }\right ) \exp\left(-a_1^2/(4q)\right) \exp\left(-a_2^2/(4q)\right) 
\label{vislattime2}
 \end{eqnarray}
Here, for an zenith scan,  $a_1 = [u-\frac{r_\nu}{2\pi}(k_{\perp 1}+k_{\perp 2}\sin\phi dH)]$ 
and $a_2 = [v-\frac{r_\nu}{2\pi}(k_{\perp 2}-k_{\perp 1}\sin\phi dH]$ 
and $q = ({1 \over \theta_0^2} - iw\pi)$. Eq.~(\ref{vislattime2}) shows that
the main impact of the w-term is to make the primary beam term complex. The 
w-term results in the information being distributed differently between
the real and imaginary part of the visibility. If we consider just the 
real part of the visibility, the primary beam appears to shrink by  a 
factor: $1/(1+\pi^2 w^2 \theta_0^4)$, which is indicative of the well-known
result that the presence of w-term decreases the angular area that can 
be  imaged. 

 The visibility correlation  is computed to be: 
\begin{eqnarray}
  \langle V_\nu(u,v,w)V^*_{\nu'}(u',v',w';t)\rangle &=& \bar{I_\nu}^2\int \frac{d^3k}{(2\pi)^3}P_{HI}(k)e^{ik_\parallel \Delta r_\nu}
  \exp\left(-ir_{\nu'} k_{\perp 1}\cos(\phi) dH\right) \nonumber \\
&\times & \left ({\pi \over p}\right )\left ({\pi \over p'}\right )  \exp\left(-{a_1^2 \over 4p}\right) \exp\left(-{a_2^2 \over 4p}\right) \nonumber \\
&\times &\exp\left(-{a_3^2 \over 4p'}\right) \exp\left(-{a_4^2 \over 4p'}\right) 
\label{viscorrfin4}
\end{eqnarray}
Here $a_3 = [u'-\frac{r_\nu}{2\pi}k_{\perp 1}]$, $a_4 = [v'-\frac{r_\nu}{2\pi}k_{\perp 2}]$, $p = (1/\theta_0^2+\theta_0^2 w^2 \pi^2)$, $p' = (1/\theta_0^2+\theta_0^2 w'^2 \pi^2)$. For $w, w'=0$, Eq.~(\ref{viscorrfin4}) reduces to Eq.~(\ref{viscorrfin1}) for  a Gaussian beam. One of the important conclusions of 
Eq.~(\ref{viscorrfin4}) is that the inclusion of w-term doesn't alter the 
nature of coherence of visibilities over time. The main impact of the w-term
 is to effectively shrink the size of primary beam from $\theta_0^2$ to $1/p$.
It can be shown that the visibility correlation scales as the primary beam
(e.g. Eqs~(11)--(13) of  \cite{2001JApA...22..293B}), and therefore, for 
non-zero $w$,  the correlation of raw visibilities results in a decrease in
 the signal.  
We note that for near coplanar array such as MWA, this effect is negligible 
for zenith drift scans. 

An important application of Eq.~(\ref{viscorrfin4}) occurs in computing the 
sensitivity of the detection of the HI signal in the tracking mode 
 ($dH = 0$ for the tracking case). As described in section~4.1, we assume 
all the visibilities 
in a narrow range of baselines to be coherent. However, these visibilities 
are computed at different times while tracking a region and therefore
 correspond to different values of $w$. Eq.~(\ref{viscorrfin4}) allows
us to compute the loss  of this correlation. 

The impact of w-term can be tackled using well-known algorithms based on
facet imaging or w-projection (for details see e.g. \cite{2008ISTSP...2..647C}).
 In other words, if raw visibilities are correlated
then we expect a small loss of signal. However, if the raw visibilities 
are first treated  by facet imaging  then the impact of w-term 
can be reduced  for either drift scans and tracking. We hope to return to this
issue in future work.

Throughout this paper we assume the primary beam to be given by Eq.~(\ref{pribeam}).  As noted above, this assumption is only valid for a  phase 
center  fixed to the zenith at the location of MWA. If the phase center is 
moved to a point on the sky that makes an angle $\delta$ with the zenith then
the projected area in that direction scales as $\cos\delta$ and the primary
beam scales as $1/cos\delta$. As noted above the HI signal 
scales as the primary beam.  The antenna gain $K$ (Eq.~(\ref{NN})) scales as the effective area of the 
telescope  or as the inverse of the primary beam. As the error on the HI 
visibility correlation scales as the square of the antenna gain (Eq.~(\ref{NN})), the signal-to-noise for the detection of the HI signal degrades as $\propto \cos\delta$. 
For instance, an equatorial drift scan would result in a loss of a factor of roughly 1.2 in signal-to-noise as compared to the zenith scan. This loss of 
sensitivity  is severer for the tracking case if regions far away from the zenith are tracked. We note that our conclusions based on the cases considered
in this paper are not altered by this loss. 

\end{appendix}

\end{document}